\RequirePackage{fix-cm}
\documentclass[twocolumn]{svjour3}          
\smartqed  
\usepackage{amsmath,amsfonts}
\usepackage{algorithmic}
\usepackage[ruled]{algorithm2e} %
\usepackage{array}
\usepackage[caption=false,font=footnotesize]{subfig}
\usepackage{textcomp}
\usepackage{stfloats}
\usepackage{url}
\usepackage{verbatim}
\usepackage{graphicx}
\usepackage[nocompress]{cite}
\usepackage{multirow}
\usepackage{diagbox}
\usepackage{tablefootnote}
\usepackage{xcolor}
\usepackage{makecell}
\usepackage{flushend} 
\usepackage[
    pagebackref=false,
    breaklinks=true,
    colorlinks=true,
    bookmarks=false,
    linkcolor=black,
    anchorcolor=black,
    citecolor=black
]{hyperref}

\newcommand{\redhighlight}[1]{\textcolor{black}{#1}}

\begin{document}

\title{Scene Graph Lossless Compression with Adaptive Prediction for Objects and Relations}

\author{Yufeng~Zhang\and
Weiyao~Lin*\and
Wenrui~Dai\and
Huabin~Liu \and
Hongkai~Xiong
}


\institute{
    Yufeng Zhang \at
      Department of Electronic Engineering, Shanghai Jiao Tong University, Shanghai China \\
      \email{worldlife@sjtu.edu.cn}
    \and
    Weiyao Lin \at
      Department of Electronic Engineering, Shanghai Jiao Tong University, Shanghai China \\
      \email{wylin@sjtu.edu.cn}
    \and
    Wenrui Dai \at
       Department of Electronic Engineering, Shanghai Jiao Tong University, Shanghai China \\
       \email{daiwenrui@sjtu.edu.cn}
    \and
    Huabin Liu \at
       Department of Electronic Engineering, Shanghai Jiao Tong University, Shanghai China \\
       \email{huabinliu@sjtu.edu.cn}
    \and
    Hongkai Xiong \at
        Department of Electronic Engineering, Shanghai Jiao Tong University, Shanghai China \\
        \email{xionghongkai@sjtu.edu.cn}
    \and
    *(Corresponding author: Weiyao Lin.)
}

\date{Received: date / Accepted: date}

\maketitle

\begin{abstract}
  The scene graph is a new data structure describing objects and their pairwise relationship within image scenes. As the size of scene graph in vision applications grows, how to losslessly and efficiently store such data on disks or transmit over the network becomes an inevitable problem.
  However, the compression of scene graph is seldom studied before because of the complicated data structures and distributions. Existing solutions usually involve general-purpose compressors or graph structure compression methods, which is weak at reducing redundancy for scene graph data. 
  This paper introduces a new lossless compression framework with adaptive predictors for joint compression of objects and relations in scene graph data. The proposed framework consists of a unified prior extractor and specialized element predictors to adapt for different data elements.
  Furthermore, to exploit the context information within and between graph elements, Graph Context Convolution is proposed to support different graph context modeling schemes for different graph elements. 
  Finally, a learned distribution model is devised to predict numerical data under complicated conditional constraints.
  Experiments conducted on labeled or generated scene graphs proves the effectiveness of the proposed framework in scene graph lossless compression task.
\end{abstract}

\section{Introduction}
\label{sec:introduction}

\begin{figure}[!t]
  \centering
  \includegraphics[width=1\linewidth]{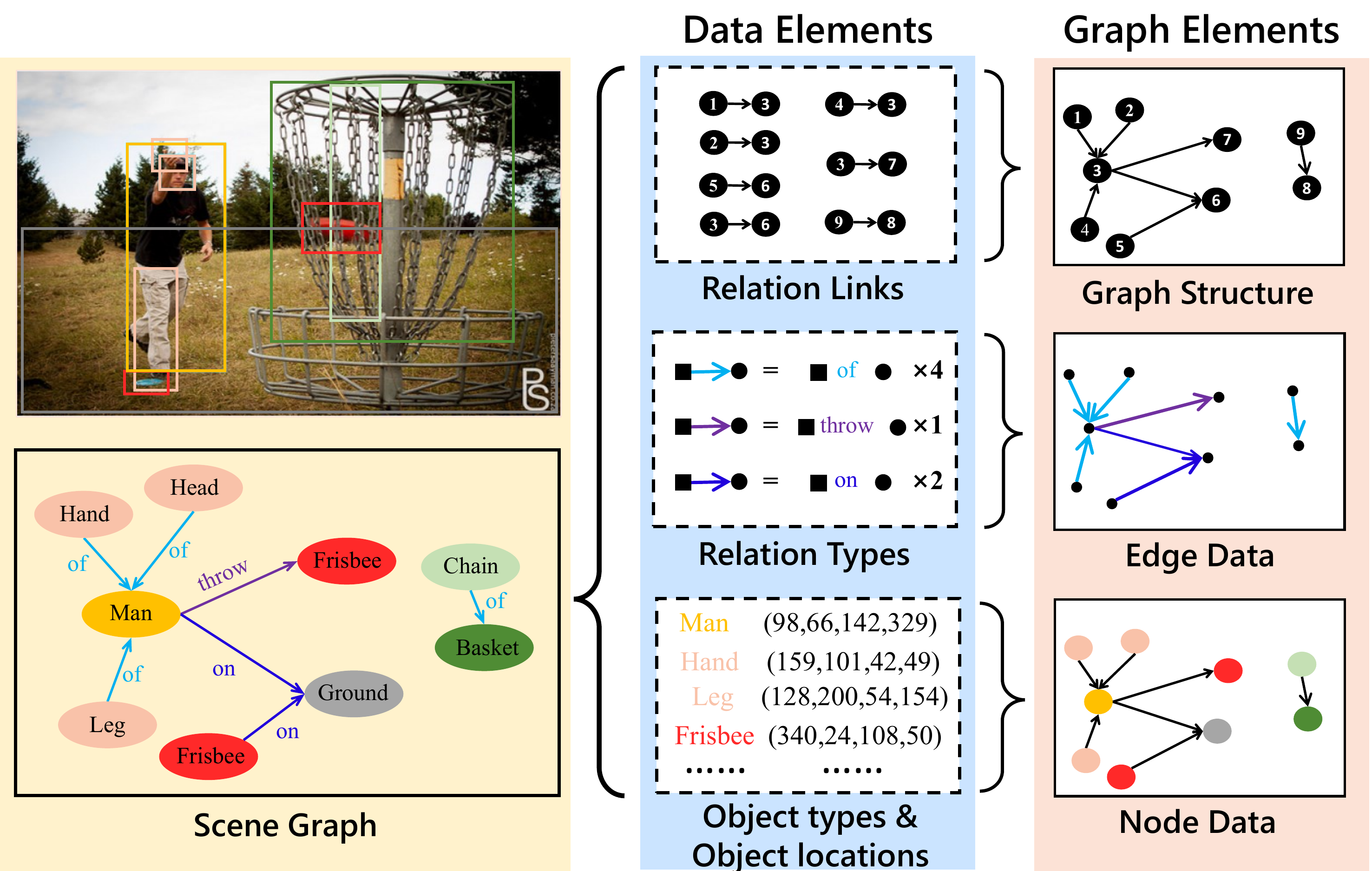}
  \caption{Scene graph example from VisualGenome\cite{DBLP:journals/ijcv/KrishnaZGJHKCKL17}. The example data has 7 relation links that forms a graph, containing 9 nodes with 9 corresponding groups of object locations and types data, as well as 7 edges with 7 corresponding relation types data.}\label{fig:dataexample}
\end{figure}

With the rapid development of computer vision, the emerging high-level visual tasks (\textit{e.g.}, visual question answering\cite{DBLP:conf/emnlp/FukuiPYRDR16, DBLP:conf/cvpr/MarinoCP0R21}, image captioning\cite{DBLP:conf/eccv/AndersonFJG16, DBLP:journals/tomccap/JiangWH21}, high-precision image retrieval\cite{DBLP:conf/wacv/Wang0YSC20, DBLP:conf/aaai/YoonKJLHPK21} and high-level video understanding\cite{DBLP:conf/cvpr/JiK0N20}) require understanding the relationships among objects in a scene. \emph{Scene graph}, a new graph-based data structure that describes \emph{objects} and their \emph{relations} in a image scene, was leveraged to address this issue\cite{DBLP:journals/ijcv/KrishnaZGJHKCKL17}.
\autoref{fig:dataexample} presents an example of scene graph composed of four \emph{data elements}: relation links, relation types, object types and object locations. 
When organized as a scene graph, data elements are categorized into three types of \emph{graph elements}: graph structure, edge data and node data. 

Recently, with the rapid increase of image and video data in both data size and content complexity, their corresponding scene graphs' data size and graph scale also increase dramatically. 
As a result, a massive volume of scene graphs needs to be stored in disks or transmitted over the network, calling for compression of scene graphs.
Moreover, lossless compression of scene graphs is necessarily required to precisely describe the relationships in scenes. 
However, this requirement is challenging for existing lossless compressors. 
The main reasons are two-fold. \textbf{First}, a scene graph contains diverse types of data elements with various shapes and distributions, as shown in \autoref{fig:dataexample}.
As far as we know, there lacks a unified system to effectively compress all of these data elements at the same time. 
\textbf{Second}, objects and relations in scene graph are closely correlated. For example, the relation `\emph{throw}' usually describes the object `\emph{man}' and a throwable object, e.g., `\emph{frisbee}'.
These conditional correlations shall be exploited to reduce the redundancy among objects and relations in scene graphs during lossless compression. 

For existing solutions, a naive solution employs a general-purpose lossless compressor like zlib~\cite{zlib} and LZMA~\cite{LZMA} to directly compress the scene graph into a bitstream. Nevertheless, such a solution ignores different shapes and distributions of data elements and correlations between objects and relations in scene graphs.
Another possible solution is applying specialized methods for different data elements, such as using WebGraph~\cite{DBLP:conf/www/BoldiV04} for compressing graph structures. 
Although this solution considers the data shape and distribution of the graph structure, it still fails to reduce potential redundancy among different graph elements.
In summary, existing solution cannot solve the above two challenges in lossless compression of scene graphs.


\begin{figure}[!t]
  \centering
   \subfloat[]{\includegraphics[width=.5\linewidth]{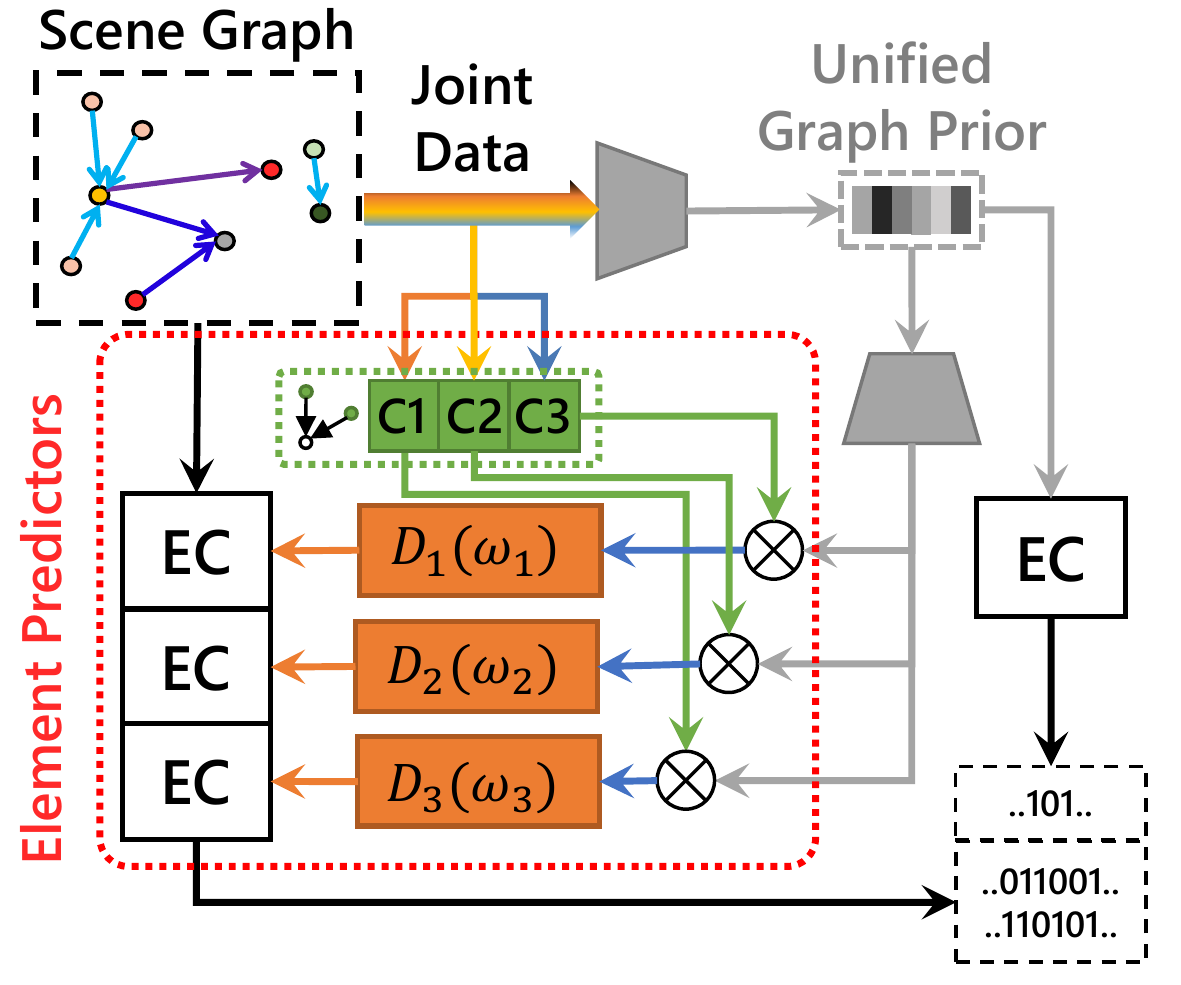}\label{fig:showcase-framework}}
   \subfloat[]{\includegraphics[width=.5\linewidth]{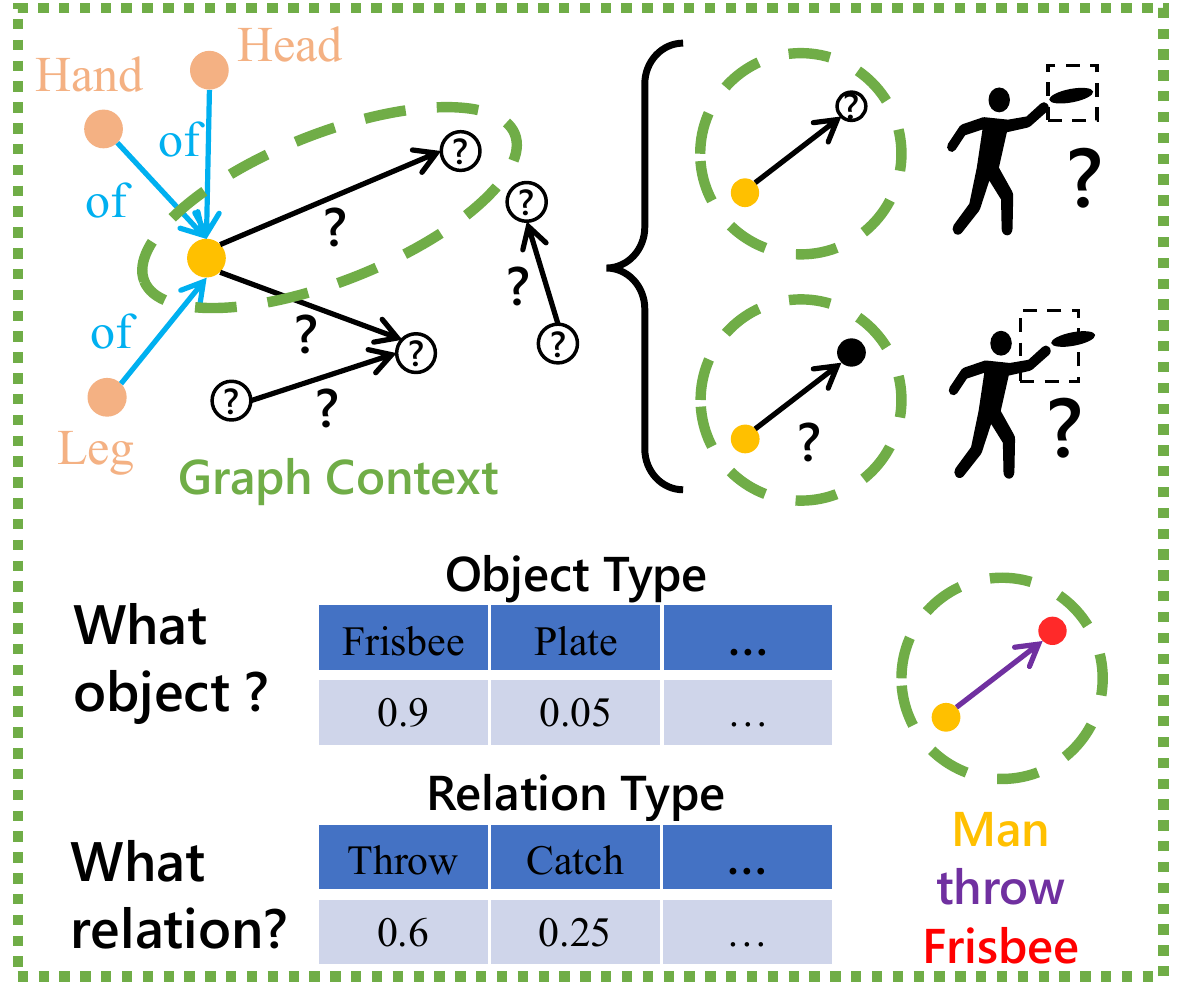}\label{fig:showcase-context}}
  \caption{(a) The proposed framework for scene graph lossless compression, EC represents the entropy coding module. The main differences of the proposed framework from the image compression framework are labelled in dotted boxes, namely element predictors (red dotted box) and graph context models (green dotted box). (b) An example of graph context modeling, one of the major contributions of this framework, is illustrated on the bottom side in the green dotted box.}
\end{figure}

To address the above challenges, a lossless compression framework for scene graphs should (1) adaptively compress different data elements in scene graphs, and (2) reduce context redundancy among objects and relations. In this paper, we consider these two requirements and propose a novel framework to achieve better lossless compression for scene graphs.

\textbf{Element-adaptive Framework} 
To this end, we propose a lossless framework that handles diverse data elements adaptively.
This is achieved by designing different \textbf{\emph{element predictors}} for distinct data elements, which learn to predict the symbol-level probabilities for each data element according to their intrinsic attributes (data shape and distribution). 
It empowers our framework to simultaneously process objects and relations in scene graphs, and meanwhile could be trained end-to-end.
\autoref{fig:showcase-framework} illustrates a sketch of our proposed framework. 
Specifically, an autoencoder-based network first extracts unified graph prior of the whole graph, providing vital information for prediction. Then, the element predictors perform adaptive predictions combining unified graph prior and various forms of context information, and convert the prediction to probabilities using numerical or categorical forms of distribution modules.

\textbf{Graph Context Models}
We further equip our element predictors with element-specific graph context models, in order to enable this framework to reduce the heavy context redundancy in scene graphs.
As presented in \autoref{fig:showcase-context}, the distribution of data elements in graph structure is closely related with their semantic relationships. For example, as the object `\emph{frisbee}' is usually used by a human, this object has a higher probability of occurring at the neighborhood of object `\emph{man}'. Similarly, knowing about object `\emph{frisbee}' and `\emph{man}', their relation type is more likely to be `\emph{throw}'. 
Graph convolution\cite{DBLP:conf/iclr/KipfW17}, which passes messages among nodes to obtain enhanced node features, is a popular solution to model such relationships. However, the element predictors in lossless compression  require that the context models should be \emph{causal} systems. In the case of scene graph data, it means that messages can only pass from the former nodes to latter nodes during context modeling. 
To address this issue, we propose a \emph{Graph Context Convolution} (GCC), which conducts message passing following the causal flows. Based on GCC, we then devise element-specific context modeling schemes for the node data predictor, graph structure predictor, and edge data predictor, respectively. They empower our framework to reduce context redundancy for all graph elements in scene graphs.

\textbf{Learned Distribution Model}
In addition, regarding \emph{numerical} data (\emph{e.g.}, object location), it is typical for element predictors to utilize a pre-defined distribution model that fits their distribution to obtain the final probabilities.
However, their distributions tend to be constrained by complicated relationships among objects in scene graphs. 
As shown in \autoref{fig:showcase-context}, the location of object `\emph{man}' could be estimated with the knowledge `\emph{head of man}', `\emph{hand of man}' and `\emph{leg of man}', where the relation `\emph{of}' indicates that they are more likely close to each other. 
Therefore, the location of object `\emph{man}' should be distributed near the object `\emph{head}', `\emph{hand}' and `\emph{leg}'.
Furthermore, an object may be related to multiple other objects with different types of relations, resulting in a more complicated distribution.
Therefore, a more general distribution model that could fit such complicated constraints is preferred. 
This paper proposes a learned distribution model consisting of a learnable dynamic neural network that defines a continuous distribution function.

By virtue of these designs, our proposed framework can losslessly compress diverse data elements and significantly reduce the context redundancy in scene graphs. Experiments demonstrate our framework achieves the best compression ratio compared to existing general or element-specific compression frameworks.
The main contributions of this paper are:
\begin{itemize}
  \item 
  We propose an element-adaptive framework for scene graph lossless compression, which is the first attempt in this field to our knowledge. 
  \item We devise element-specific graph context models for different graph elements, which significantly reduce the context redundancy during compression.
  \item We propose a learned distribution model to adaptively fit the complicated distribution of numerical data in scene graphs.
\end{itemize}

The remainder of this paper is organized as follows.
\autoref{sec:relatedworks} reviews related works.
\autoref{sec:overview} gives an overview of the proposed framework.
\autoref{sec:method-cm} introduces the proposed graph context models for different graph elements. 
\autoref{sec:method-dm} introduces the learned distribution model. \autoref{sec:experiments} shows experimental results on various datasets and corresponding analysis.
\autoref{sec:conclusion} finally summarizes this paper.

\section{Related Work}
\label{sec:relatedworks}
\subsection{General-purpose Lossless Compression}
General purpose lossless codecs usually convert any forms of data into a sequence of symbols, and compress the sequence into a compressed stream. 

A popular tool for general-purpose lossless compression is zlib~\cite{zlib}, which combines LZ77~\cite{DBLP:journals/tit/ZivL77} and Huffman coding~\cite{4051119Huffman}.
Many later compression codecs improve such scheme by optimizing the string matching algorithm in LZ77 (such as LZMA\cite{LZMA} and Brotli\cite{DBLP:journals/tois/AlakuijalaFFKOS19}) or switching to better entropy coder such as  Asymmetric Numeral System\cite{DBLP:journals/corr/Duda13} (such as Zstd\cite{Zstd}). 
Some other codecs designs a context-adaptive entropy coder to exploit context information. For example, PPM\cite{1096090PPM} utilize a simple suffix-based context model to count symbol frequencies adaptively, while PAQ\cite{Mahoney2005AdaptiveWO} utilize the context mixing scheme to combine predictions from many specially designed context models.
Recent works also explores using neural networks to learn prior information of data or exploit context information more robustly, such as CMIX\cite{CMIX}, and Dzip\cite{DBLP:conf/dcc/GoyalTCO20}.


However, for scene graph data, such codecs ignore hidden redundancy between different data elements. 
Furthermore, context information in general-purpose compression methods are usually explored in adjacent string of symbols, while for graph-structured data like scene graphs, adjacency usually exists along graph edges. 
Therefore, applying general-purpose codecs is an inefficient solution for scene graph lossless compression.

\subsection{Graph Structure Lossless Compression}

Graph is a common form of data in many areas, such as web-page database, social network and biological data. Therefore, graph structure lossless compression has also been studied in many previous works, as reviewed in \cite{DBLP:journals/corr/abs-1806-01799}.
For example, WebGraph\cite{DBLP:conf/www/BoldiV04} is a framework designed for web graph compression. Based on the locality and sparsity features of web-page URLs, WebGraph utilizes multiple compression techniques including reference compression and differential compression to exploit such features.
Another compression method built on such features was $k^2$-tree\cite{DBLP:conf/spire/BrisaboaLN09}, which iteratively divides the graph adjacency matrix into $k^2$ sub-matrices, and stops when the sub-matrix is a all-zero matrix. 


Although such methods could be applied for graph structure compression, the compression of node data and edge data is usually ignored. In this work, in addition to graph structure compression, node data compression and edge data compression should also be considered.

\subsection{Learning-Based Image Compression}
Recently, many compression related works has been using neural networks.
A large portion of those works are related to image compression.
Most recent approaches of deep-learning based image compression utilize an autoencoder-based end-to-end framework \cite{DBLP:conf/iclr/BalleLS17}.
Following \cite{DBLP:conf/iclr/BalleLS17}, \cite{DBLP:conf/iclr/BalleMSHJ18} proposed a hyperprior autoencoder that enables better latent prediction. 
Afterwards, a few works propose to improve the hyperprior model by exploiting context information from neighbouring pixels  (\cite{DBLP:conf/nips/MinnenBT18,Chen9359473}) or improving the distribution module with more complex distributions (\cite{DBLP:conf/cvpr/ChengSTK20,DBLP:journals/corr/abs-2107-06463}). 
Other frameworks also explores different network structures such as context-based convolutional networks for entropy modeling\cite{Li9067005} or wavelet-like convolutional networks for image transformation\cite{DBLP:journals/tmm/Ma0X020}.

For lossless or near-lossless image compression, a straightforward solution is to apply a predictor and compress the residual~\cite{9277919}.
Also, some works apply autoencoder as likelihood generative networks~\cite{DBLP:conf/cvpr/MentzerATTG19,DBLP:conf/icml/KingmaAH19} to estimate pixel-wise probability for entropy coding in order to achieve lossless compression. Other works deploy invertable neural networks~\cite{DBLP:conf/nips/HoogeboomPBW19, 9204799} to generate a lossless transformation of the original data for easier entropy coding.

The proposed framework is inspired by many successful learning-based compression frameworks~\cite{DBLP:conf/iclr/BalleMSHJ18,DBLP:conf/nips/MinnenBT18, DBLP:conf/cvpr/MentzerATTG19, DBLP:conf/icml/KingmaAH19}. However, the proposed framework is specially designed for compression of the complete scene graph. 
To achieve this, some novel modules are proposed and designed, such as the adaptive element predictors with graph context models.

\section{Proposed Framework} 
\label{sec:overview}

\begin{figure*}[!t]
  \centering
  \includegraphics[width=1.0\textwidth]{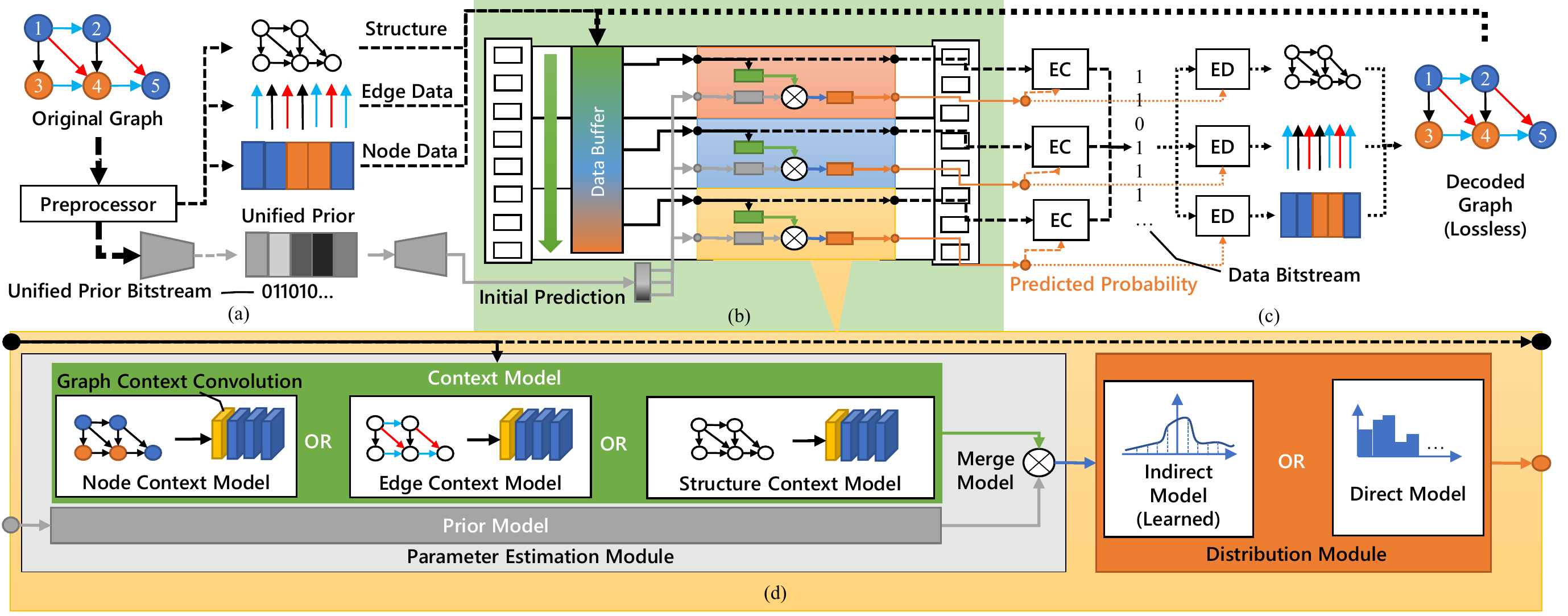}
  \caption{
  Overview of the proposed framework. (a) Unified Prior Extractor (b)  Element Predictors (c) Entropy Coding (d) Detailed overview of an element predictor.
The lines between modules indicate data flows: (1) the dashed lines means data flows that only exists on the compressor side, (2) the dotted lines means data flows that only exists on the decompressor side, and (3) the solid lines means data flows that exists on both sides.
EC and ED represents entropy coding and entropy decoding, respectively.
  Best viewed in color.
 }
  \label{fig:overview}
\end{figure*}
\autoref{fig:overview} illustrates the proposed framework that consists of three essential parts: i) the unified prior extractor (\autoref{fig:overview}\redhighlight{a}), ii) the element predictors (\autoref{fig:overview}\redhighlight{b}), and iii) the entropy coders (\autoref{fig:overview}\redhighlight{c}). The unified prior extractor aims to extract critical information from the whole scene graph to provide a unified prior for subsequent compression. Then, an initial prediction for the original graph data could be produced with the unified prior. 
The second part comprises multiple element-adaptive predictors, designed according to each data element's attributes (data shape \& distribution) to predict symbol-wise probabilities for the entropy coder. Specifically, each element predictor consists of two main modules: the parameter estimation module and distribution module, as depicted in \autoref{fig:overview}\redhighlight{d}. 
For the inputs of element predictors, the data buffer stores the original data (in compression) or the losslessly decompressed data (in decompression). Meanwhile, this buffer decides the order of data elements. Therefore, it also controls the order and structure of element predictors. 
Finally, the lossless entropy coders leverage the predicted probability to perform lossless compression and decompression on the data elements. We will elaborate on them in the following sections.

In the rest of this paper, we use normal font for scalar variables and functions, bold lowercase font for vector variables, and bold uppercase font for matrices. We reserve $tril(\mathbf{A})$ to represent the lower triangle part of a matrix $\mathbf{A}$.
To avoid confusion, we use the term ``decoder'' to describe the generator module of the autoencoder and the term ``decompressor'' to describe the module in a compression framework that converts a compressed bitstream back to the original data.

\subsection{Unified Prior Extractor}
\label{sec:overview-pcm}
The unified prior extractor is expected to provide a coarse perception of the entire graph in the compressed bitstream for element predictors. We implement it with the popular autoencoder strcture. Specifically, the encoder first takes the entire preprocessed graph as input and produces a latent vector as the prior information. 
Then, the decoder receives the prior information and outputs an initial prediction that is fed into the element predictors to participate in the final prediction of symbol probability for lossless entropy coding. The above process is illustrated in \autoref{fig:overview}\redhighlight{a}.


Furthermore, as the latent vector should be finally compressed into a bitstream, we need to calculate the entropy of the latent vector and utilize it to estimate and optimize the length of compressed unified prior bitstream during training. For latent vector $\mathbf{y}$, its entropy is estimated by:
\begin{eqnarray}\label{eq:loss-latent}
  \mathcal{H}_{p} = \mathbb{E}_{\mathbf{y} \sim P} (-\log_{2}P(\mathbf{y}))
\end{eqnarray}
where $P(\mathbf{y})$ is the probability of latent vector $\mathbf{y}$.
To obtain $P(\mathbf{y})$ 
, we adopt the same factorized-prior model as \cite{DBLP:conf/iclr/BalleMSHJ18}. It learns a fixed factorized prior distribution to fit the latent vector's distribution during training.


\subsection{Element Predictors}
\label{sec:overview-ppem}
In scene graphs, the data shapes, contextual relationships, and distributions vary from data elements. Based on this observation, we propose the element-adaptive predictors for scene graph lossless compression, which empower our framework to cope with diverse data elements adaptively in scene graphs.
Specifically, each predictor consists of two modules: the \textbf{parameter estimation module} and the \textbf{distribution module}. The former obtains element-specific information according to data shape and contextual relationship, while the latter fits different data distributions of data elements based on the element-specific information.

Specifically, the parameter estimation module first merges the context information extracted from the data buffer and the initial prediction to have a complete perception of data elements. Then, it can estimate distribution parameters for data elements according to the perception. To this end, we implement parameter estimation module with three sub-models to accomplish the above process: the context model, the prior model, and the merge model. Note that we design an element-specific context model for each graph element. As shown in \autoref{fig:overview}\redhighlight{d}, it has basically three options in scene graphs: the node context model, edge context model, and graph structure context model.
Detailed algorithms and implementations of the parameter estimation module, especially the context model, will be discussed in \autoref{sec:method-cm}.

The distribution module contains a random distribution model to fit different data distributions. It receives the estimated parameters given by the parameter estimation module and predicts symbol-wise probabilities. As shown in \autoref{fig:overview}\redhighlight{d}, two forms of distribution models are available for different data distribution. One is a direct distribution model for categorical data such as object and relation types. 
Another is an indirect distribution model for numerical data such as object locations. Moreover, we propose a learned distribution model to tackle the complicated distribution of numerical data.  
Details will be discussed in \autoref{sec:method-dm}.

The calculation process of the element predictor could be represented as \autoref{eq:element-predictor}.
\begin{eqnarray}
\label{eq:element-predictor}
  P(\mathbf{x}) = f_{D}(\mathbf{x}, \mathbf{\omega})
\end{eqnarray}
where $f_{D}$ represents the distribution model, $\mathbf{x}$ represents the data element, and $\mathbf{\omega}$ represents distribution parameters calculated by the parameter estimation module, represented as \autoref{eq:pem}.
\begin{eqnarray}\label{eq:pem}
  \mathbf{\omega} = f_{PEM}(\mathbf{x}, \mathbf{y}) = f_{M}(f_{C}(\mathbf{x}), f_{P}(\mathbf{y}))
\end{eqnarray}
$f_{PEM}$ is the parameter estimation module, and $ f_{P} $,$ f_{C}$,$ f_{M} $ are its prior model, context model and merge model respectively.

Note that there may exist other data elements than those 4 data elements in \autoref{fig:dataexample} in different scene graphs. For example, the "relation weights", namely the certainty level of the estimated relation, and the "human pose", namely the skeleton points on a human object. 
The proposed element predictor can generalize to various forms of data elements in scene graphs by reusing the proposed parameter estimation modules and distribution modules. 
An example of the implementation of the element predictor for some possible data elements in scene graph is described in \autoref{tab:impl}.
For example, object locations in scene graph belong to node data in graph, and is numerical data. Therefore, the element predictor for object locations should apply node context model and indirect distribution model. Relation types belong to edge data in graph, and is categorical data. Therefore, the element predictor for relation types should apply edge context model and direct distribution model.
Implementations for other data elements such as object type and relation weight could also be inferred accordingly.

\begin{table}[!t]
  \centering
  \scriptsize
  \setlength{\tabcolsep}{4pt}
  \caption{Implementation of the element predictor for different elements in scene graph.
}
  \label{tab:impl}
\begin{minipage}{1.\linewidth}
  \begin{tabular}{lllll}
  \hline
  Data Element & Graph Element & $f_{PEM}$      & $f_{D}$    \\
  \hline
  Object Location     & Node Data & Node Context & Indirect  \\
  Object Type        & Node Data & Node Context & Direct   \\
  Relation Link       & Structure      & Structure Context    & Direct   \\
  Relation Type      & Edge Data & Edge Context  & Direct   \\
  Relation Weight  & Edge Data & Edge Context  & Direct  \\
  \hline           
  \end{tabular}
\end{minipage}
\end{table}

\subsection{Entropy Coding}
Lossless coding for a scene graph is achieved by directly applying entropy coders on the scene graph data with the probabilities predicted by element predictors. The whole coding is free from any processes that may introduce data loss, such as quantization.
Note that although quantization is applied in our unified prior extractor, its initial prediction output is used for probability prediction, and thus it will not affect the losslessness of the original data. 
Moreover, the preprocessor shown in \autoref{fig:overview}\redhighlight{a} only performs node sorting for better context modeling and will not introduce any data loss. More details about this issue will be discussed in \autoref{sec:method-cm}.

During training, the predicted probability is used to calculate the data entropy to estimate the length of the compressed data bitstream according to \autoref{eq:loss-data}.
\begin{eqnarray}\label{eq:loss-data}
  \mathcal{H}_{d} = \mathbb{E}_{\mathbf{x} \sim P} (-\log_{2}P(\mathbf{x}))
\end{eqnarray}
where $P(\mathbf{x})$ is provided by the element predictor as in \autoref{eq:element-predictor}
.

\subsection{Optimization}
\label{sec:overview-loss}
For entropy coding, the entropy value indicates the length of compressed data. Consequently, the entropy of the unified prior and all data elements should be utilized to calculate the training loss, which is formulated as Eq. \ref{eq:loss}:
\begin{eqnarray}\label{eq:loss}
  \mathcal{L} = \frac{\mathcal{H}_{p} + \sum_{i} \mathcal{H}_{d,i}}{N} 
\end{eqnarray}
where $\mathcal{H}_{d,i}$ is the entropy value of the $i$-th data element. $N$ is the totoal number of nodes, which is used for normalization. The above loss function encourages our framework to reduce the entropy, \emph{i.e.}, the compression ratio.

\section{Context Modeling for Graph Elements}
\label{sec:method-cm}
In this section, the parameter estimation modules in predictors for different graph elements are discussed.
The main difference among graph elements (graph structure, edge \& node data) is the data shape, which decides how to gather the context information. 
Thus, we will focus on the context models in parameter estimation. 

On the decompressor side, the context model uses decompressed information to estimate the probability of current data. Importantly, to ensure a correct entropy decoding, the compressor should aggregate the same information as the decompressor to produce the same probability. This requirement could be called the `\textbf{\textit{Context Condition}}.' To make all graph elements in scene graphs satisfy this condition, we devise different context models for their corresponding predictors. The following sections will elaborate on proposed context modeling approaches for all the three graph elements (node data, graph structure, and edge data).

\subsection{Graph Context Convolution for Node Data}
\label{sec:method-cm-gcc}
We begin by devising the context model for the node data predictor. To satisfy the context condition, the node data predictor should be a causal system. Specifically, when predicting any node, we can aggregate information from previous nodes to predict the current one. On the other hand, information from latter nodes should be excluded from aggregation. Based on the graph convolution, we devise the Graph Context Convolution (GCC), which can well aggregate context information in node data while satisfying the context condition.

Vanilla graph convolution passes messages according to the adjacency matrix of the graph. A straightforward revision to satisfy the context condition is to cut the message passing flow from the latter to former nodes. In this way, we can obtain a causal message passing flow. To this end, the solution for GCC is proposed by removing the lower triangle part of the adjacency matrix before performing message passing, represented by \autoref{eq:gcc}:
\begin{eqnarray}
\label{eq:gcc}
\mathbf{X_{n+1}} = H(\mathbf{\hat{A}}) \mathbf{X_{n}} \mathbf{\theta},~\mathbf{\hat{A}} = \mathbf{A} - tril(\mathbf{A})
\end{eqnarray}
where $\mathbf{X_{n}}$ is node features at layer n, and $H$ is the function that generates the message passing matrix, which varies with different graph convolution operations. $\mathbf{\hat{A}}$ is the adjacency matrix removing the lower triangle part, and $\mathbf{\theta}$ denotes the convolution kernel parameters. 
Note that any self-loop in the message passing matrix $H(\mathbf{\hat{A}})$ should also be removed since the decompressor cannot gather information from the node itself when decoding.
Specifically, for GCN-based\cite{DBLP:conf/iclr/KipfW17} GCC operation, it can be formulated by:
\begin{eqnarray}
\label{eq:gcc-gcn}
  \mathbf{X_{n+1}} = \mathbf{D}^{-\frac{1}{2}} \mathbf{\hat{A}} \mathbf{D}^{-\frac{1}{2}} \mathbf{X_{n}} \mathbf{\theta}
\end{eqnarray}
where $\mathbf{D}$ is the degree matrix of $\mathbf{\hat{A}}$. 

\begin{figure}[!t]
  \centering
    \includegraphics[width=1\linewidth]{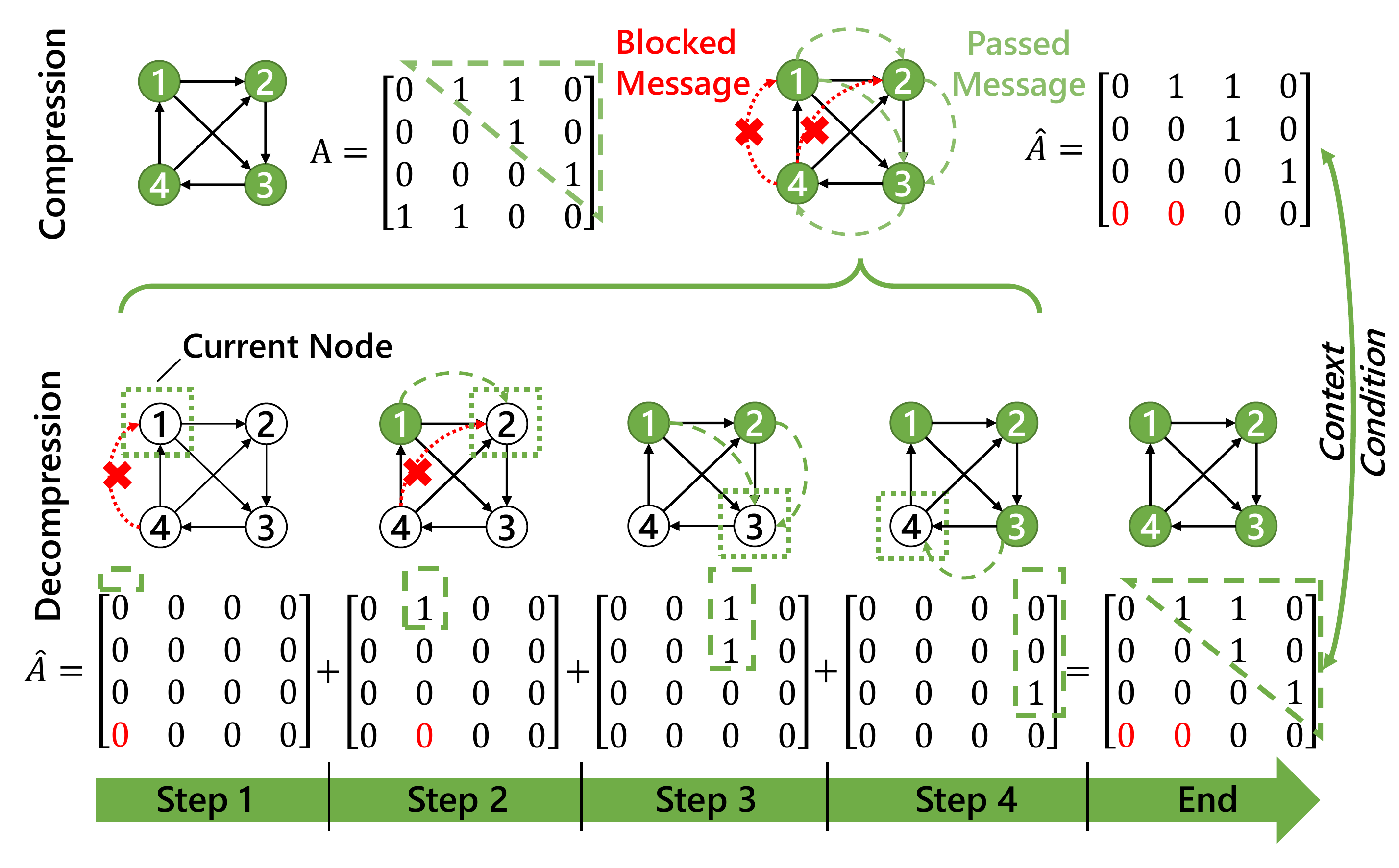}
  \caption{An example of Graph Context Convolution (GCC).}
  \label{fig:gcc-example}
\end{figure}

\autoref{fig:gcc-example} illustrates an example of the GCC for node data context modeling. During decompression, when decompressing node-1, node-4 is unknown, so $\text{edge}_{4\rightarrow1}$ is unavailable in message passing. When decompressing node-2, node-1 has already been decompressed while node-4 has still not, so $\text{edge}_{1\rightarrow2}$ is used for message passing and $\text{edge}_{4\rightarrow2}$ is unavailable. When decoding node-3\&4, all connecting edges are available. By adding all message passing matrices used in the decompressing steps, we can obtain $\mathbf{\hat{A}}$ (in \autoref{eq:gcc}) for compression.

Although the implementation of GCC is relatively simple and may introduce some context information losses in practice
, it's still a lightweight and effective operation to extract node contextual information by causal message passing.
Moreover, we further apply a pre-processor to address the issue of context information loss (details will be discussed in \autoref{sec:method-cm-igc}).

\subsection{Directed Graph Context Autoencoder for Graph Structure}
\label{sec:method-cm-dgcae}
\begin{figure}[!t]
  \centering
\vspace{-5pt}
  \subfloat[]{\includegraphics[width=.15\linewidth]{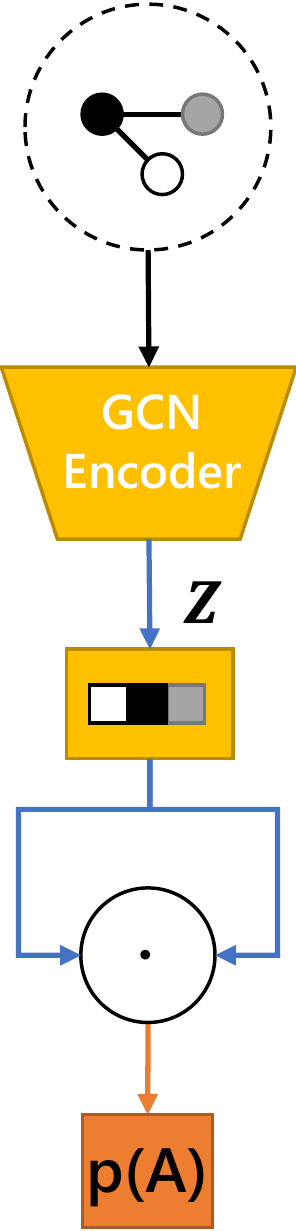}\label{fig:gae}}
  \subfloat[]{\includegraphics[width=.35\linewidth]{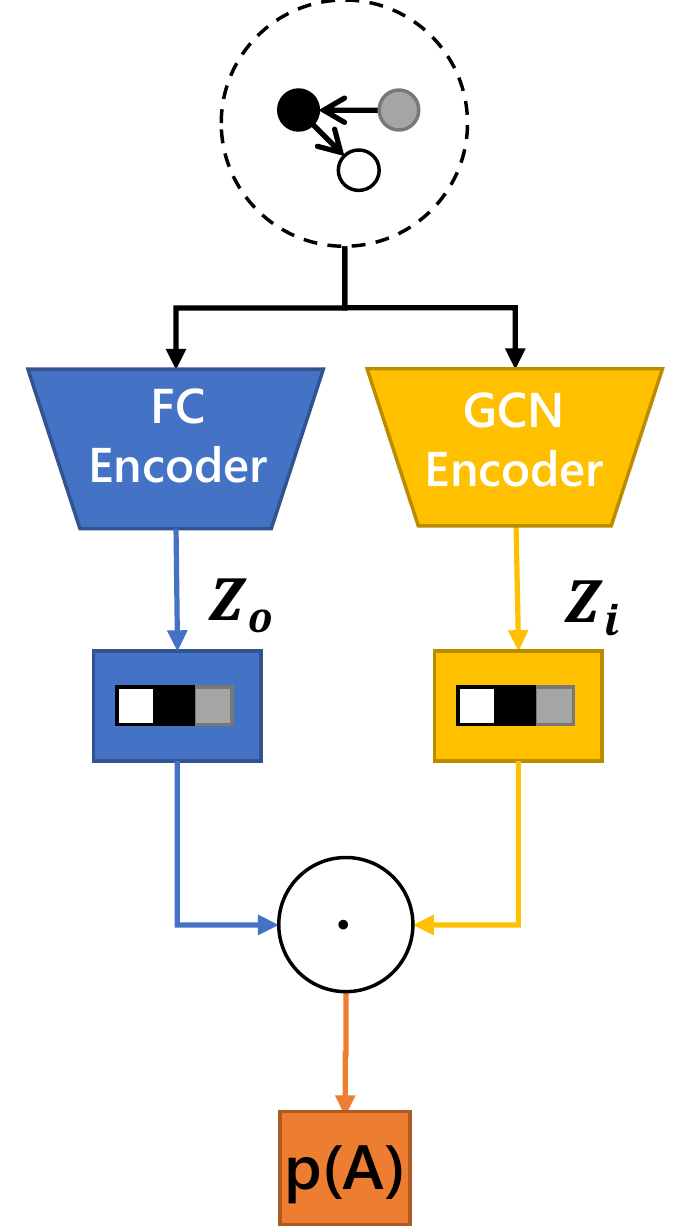}\label{fig:dgae}}
  \subfloat[]{\includegraphics[width=.5\linewidth]{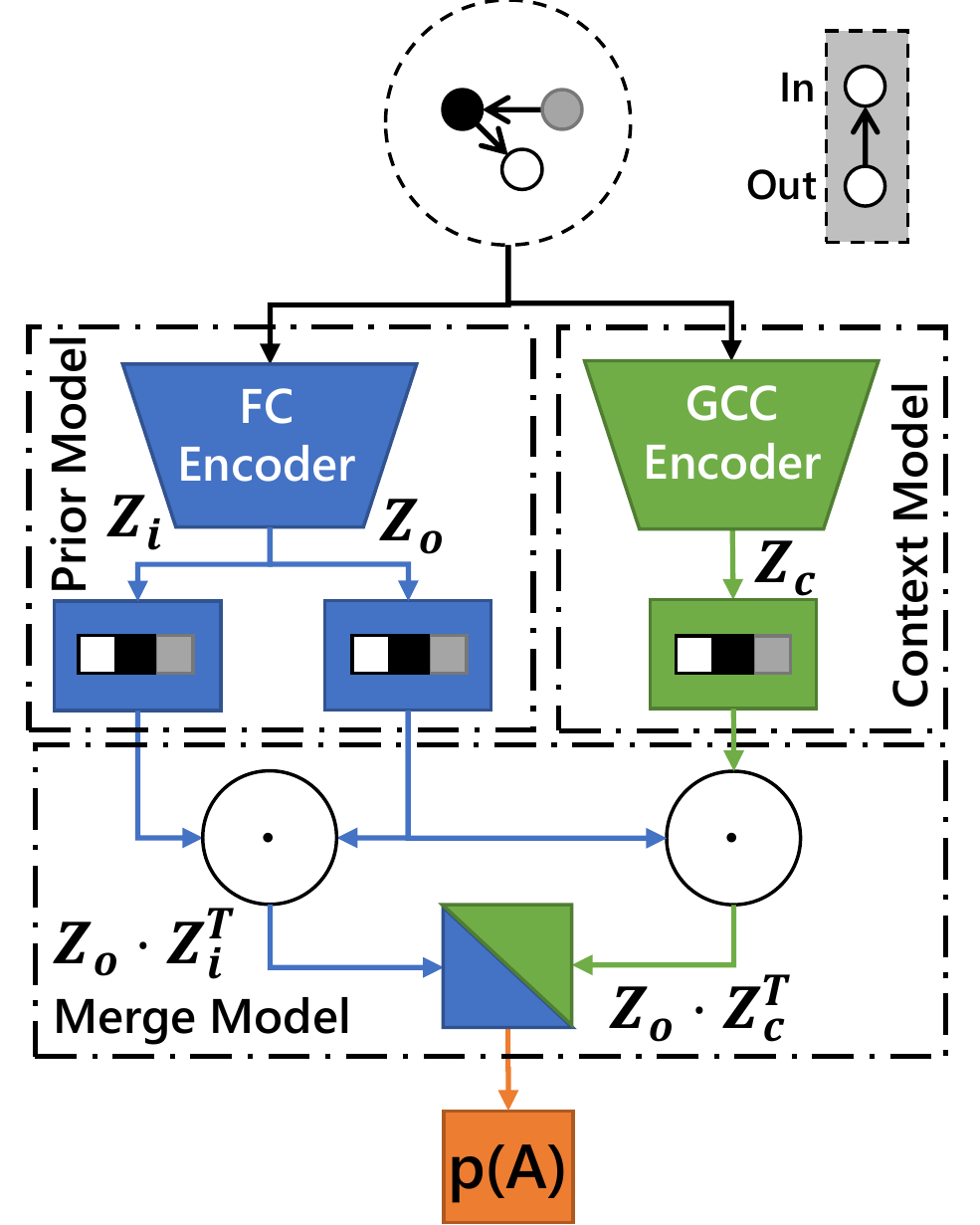}\label{fig:dgcae}}
  
   \caption{Structure of: (a) The original Graph Autoencoder (GAE)\cite{DBLP:journals/corr/KipfW16a}. (b) Directed Graph Autoencoder (DGAE). (c) Directed Graph Context Autoencoder (DGCAE).}
  \label{fig:dgcae-idea}
\end{figure}

For graph structure, existing works usually apply graph autoencoders\cite{DBLP:journals/corr/KipfW16a} to embed graph structure into a latent vector, which could be used to predict its adjacency matrix, as illustrated in \autoref{fig:gae}. However, they could only process undirected graphs, thus cannot be applied to directed graph structure in most scene graphs. Therefore, we devise the Directed Graph Context Autoencoder (DGCAE) based on GCC to embed context information for the directed graph structure in scene graphs.

First, in order to process directed graph structure in scene graph data, we design the Directed Graph Autoencoder (DGAE).
For directed graphs, the graph convolution passes messages from the out-nodes to the in-nodes along the edge direction. Therefore, the node messages 
are passed into the nodes pointed by directed edges. For this reason, we call such features obtained by node message passing as `\emph{node-in}' latent feature (denoted by $\mathbf{Z}_{i}$). 
Furthermore, in order to properly decode $\mathbf{Z}_{i}$ as an asymmetric prediction matrix for a directed graph, another `\emph{node-out}' latent feature (denoted by $\mathbf{Z}_{o}$) is then required to decode $\mathbf{Z}_{i}$ to obtain the prediction of the directed graph adjacency matrix. 
$\mathbf{Z}_{o}$ can be obtained by directly transforming the node data with an fully-connected encoder. 
The decoder can be implemented by a dot-product operation between $\mathbf{Z}_{o}$ and $\mathbf{Z}_{i}$: 
\begin{eqnarray}
  \label{eqn:dgae}
  p(\mathbf{A} \vert \mathbf{Z}_{i} , \mathbf{Z}_{o})
   = \sigma(\mathbf{Z}_{o} \cdot \mathbf{Z}_{i}^{T})
\end{eqnarray}
where $\sigma$ is the sigmoid function. The whole pipeline of DGAE is illustrated in \autoref{fig:dgae}.

Second, to model context information for directed graphs, we further propose a DGAE variant by replacing the GCN encoder in DGAE with a GCC-based encoder. In this way, the graph structure context information can be also embedded into a latent vector (denoted as $\mathbf{Z}_{c}$), replacing $\mathbf{Z}_{i}$ in DGAE. 
However, since the proposed GCC disallows the message passing from latter nodes to former nodes, $\mathbf{Z}_{c}$ can only be used to decode edges that satisfy the context condition. 
Therefore, the lower triangle part of the decoded adjacency matrix cannot utilize $\mathbf{Z}_{c}$. To deal with the lower triangle part, another autoencoder composed of fully-connected layers should be applied to obtain a pseudo node-in feature $\mathbf{Z}_{pi}$. 
By combining the matrices produced by both decoder, the final prior estimation result for the adjacency matrix are decoded. 
The decoding process could then be represented as in \autoref{eqn:dgcae}.
\begin{eqnarray}
  \label{eqn:dgcae}
  p(\mathbf{A} \vert \mathbf{Z}_{i}, \mathbf{Z}_{o}, \mathbf{Z}_{c}) = \left\{
  \begin{aligned}
    \sigma(\mathbf{Z}_{o} \cdot \mathbf{Z}_{pi}^{T}) & , & A_{jk} \in tril(\mathbf{A}). \\
    \sigma(\mathbf{Z}_{o} \cdot \mathbf{Z}_{c}^{T}) & , & otherwise, \\
  \end{aligned}
  \right. 
\end{eqnarray}
We call such variation of DGAE as Directed Graph Context Autoencoder (DGCAE), as shown in \autoref{fig:dgcae}. 

By applying DGCAE, prediction of the graph structure in the upper half of the adjacency matrix could utilize both context information and prior information, and the lower half is predicted with prior information only. Note that DGCAE already includes the context model, prior model and merge model, as illustrated with dashed boxes in \autoref{fig:dgcae}, and forms a complete parameter estimation module. 



\subsection{Edge Graph Construction for Edge Data}
\label{sec:method-cm-egc}
Since scene graphs also contain numerous edge data, an edge data context modeling approach is necessary for the edge data predictor. However, different from node data, whose shape is independent of the graph structure, the shape of edge data is closely related to the graph structure. Similarly, the proposed GCC-based predictor cannot be directly applied to edge data context modeling. A naive idea is to convert edge data to adjacency matrices and apply the proposed DGCAE to process them. However, since most scene graphs are relatively sparse, the converted matrices may occupy much more memory than the original edge data. 

\begin{figure}[!t]
  \centering
    \includegraphics[width=8cm]{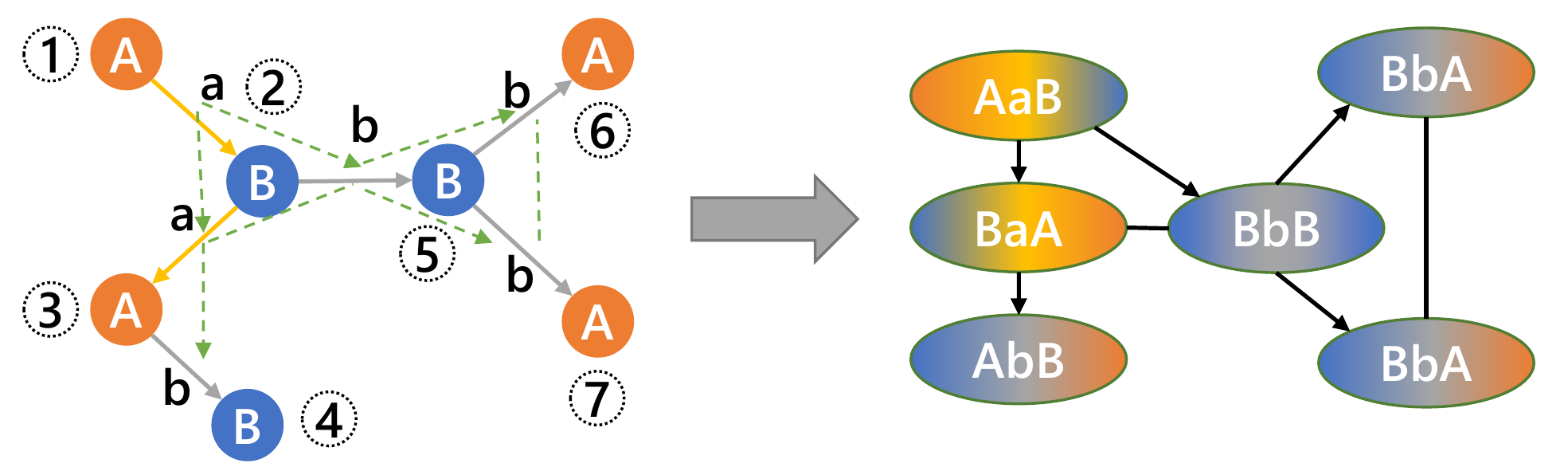}
  \caption{Examples of edge graph construction. Left: original graph. Right: graph constructed by EGC.}
  \label{fig:egc-examples}
\end{figure}

Similar to graph nodes, edges that share the same node also possess a close semantic correlation. Therefore, we can also model the context information for edge data by message passing. To this end, we propose an algorithm to convert edge data to node data, which enables the GCC-based network to model edge data context information. We call such conversion as Edge Graph Construction (EGC). Specifically, it constructs a new graph by keeping the original edge data on new nodes and adding connections between edges that share the same node.
\autoref{fig:egc-examples} illustrates an example of EGC. 
Also, the detailed process of EGC is described in \autoref{alg:egc}.

\begin{algorithm}[t]
\small
  \caption{Edge Graph Construction}
  \label{alg:egc}
  \LinesNumbered
  \KwIn{input graph $\mathbf{G}$ with nodes $ \mathbf{V} $ and edges $\mathbf{E}$}
  \KwOut{output graph $\mathbf{G}_{e}$}

  Initialize empty graph $\mathbf{G}_{e}$.

  \tcp{Step1: Node Pairing. $E(i, j, \mathbf{p})$ stands for an edge with edge data $\mathbf{p}$} connecting the i-th node $\mathbf{V}_{i}$ and j-th node $\mathbf{V}_{j}$.
  
  \ForEach{Edge $E(i, j, \mathbf{p})$ in $\mathbf{E}$}{
      AddNode($\mathbf{G}_{e}$, concatenate($\mathbf{V}_{i}, \mathbf{V}_{j}, \mathbf{p})$))
  }

  \tcp{Step2: Pairwise Edge Connection}

  \ForEach{Edge Pair $E_{k}(i_{k}, j_{k}, \mathbf{p}_{k}), E_{l}(i_{l}, j_{l}, \mathbf{p}_{l})$ in $\mathbf{E}$}{
    \If{$i_{k}$ == $i_{l}$ or $j_{k}$ == $j_{l}$}{
      AddUndirectedEdge($\mathbf{G}_{e}$, k, l)
    }
    \If{$j_{k}$ == $i_{l}$}{
      AddDirectedEdge($\mathbf{G}_{e}$, k, l)
    }
    \If{$i_{k}$ == $j_{l}$}{
      AddDirectedEdge($\mathbf{G}_{e}$, l, k)
    }
  }

  \Return{$\mathbf{G}_{e}$}
\end{algorithm}


Note that in the node pairing step, both node and edge data in the original graph can be used to construct the node data in the new graph. If node data are not available at the stage of edge data predictor, we use the unified prior information instead.
In the edge connection step, the newly constructed graph is generated according to the edge adjacency in the original graph. As shown in \autoref{fig:egc-examples}, for adjacent edges that forms a path (\emph{e.g.}, $\text{edge}_{1\rightarrow2}$ and $\text{edge}_{2\rightarrow3}$), EGC will connect them with directed edges. In contrast, adjacent edges that share the same out-node or in-node (\emph{e.g.}, $\text{edge}_{2\rightarrow3}$ and $\text{edge}_{2\rightarrow5}$ share the same out-node 2) will be connected with undirected edges.

After applying EGC, we can reuse the Graph Context Convolution (GCC) to aggregate the context information in edge data. 

\subsection{Preprocessor for Directed Graph Context Improvement}
\label{sec:method-cm-igc}
As discussed in \autoref{sec:method-cm-gcc}, the Graph Context Convolution (GCC) follows causal message passing flows. Therefore, all edges that start from latter nodes to former nodes (or non-causal edges) will be dismissed in message passing, which may introduce losses to context information. In order to address this issue for GCC-based context models, a preprocessing step is proposed to preserve more context information.

\begin{figure}[!t]
  \centering
  \subfloat[]{\includegraphics[width=1\linewidth]{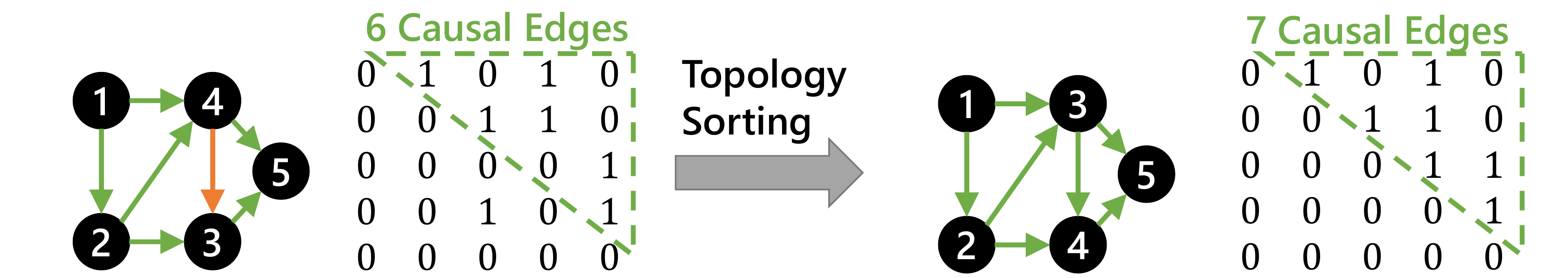}\label{fig:ci-topo}}
  \vfill
  \subfloat[]{\includegraphics[width=1\linewidth]{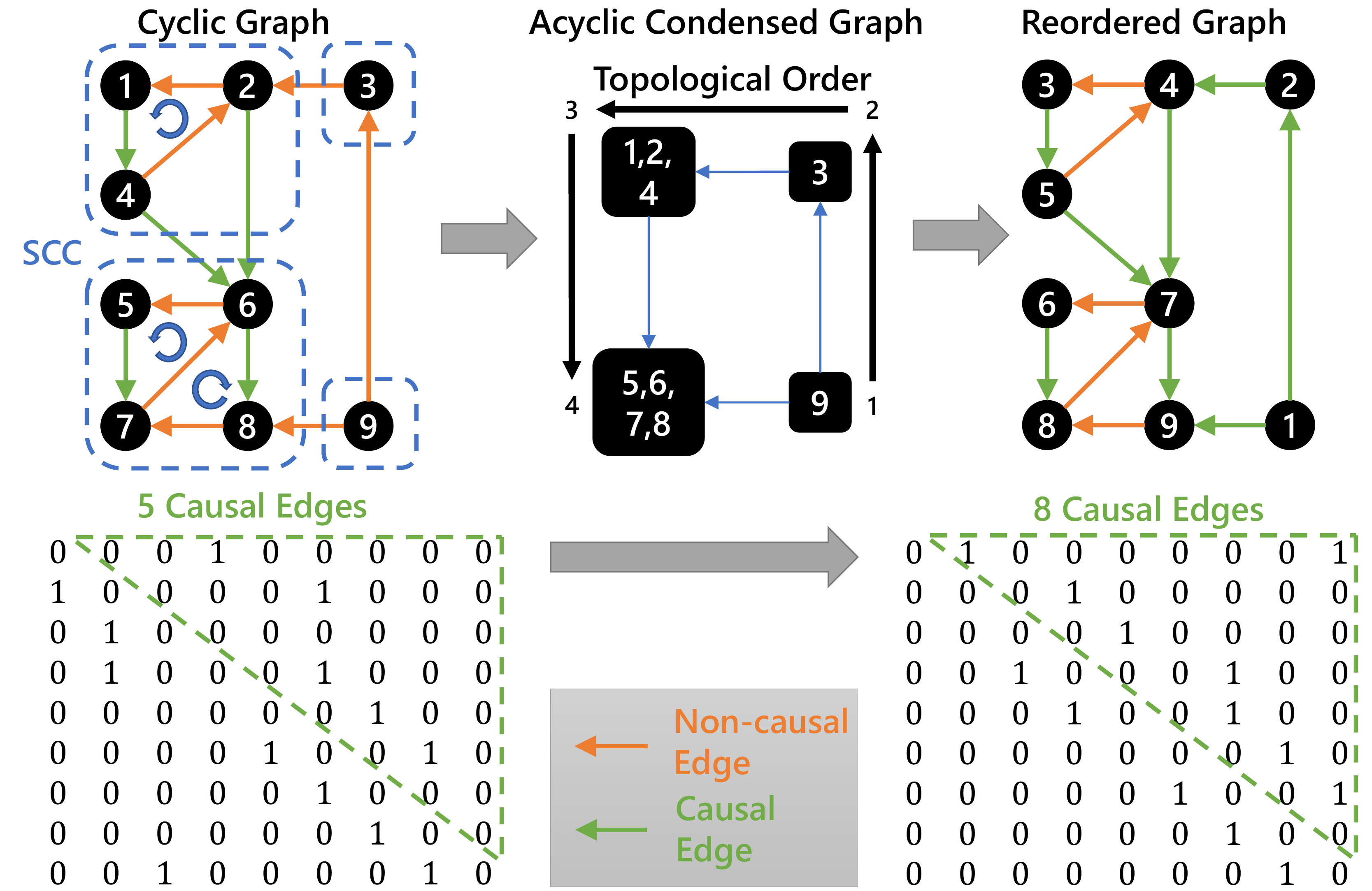}\label{fig:ci-reorder}}
  
  \caption{Context Improvement Examples: (a) Topology sorting for directed acyclic graph. (b) Proposed workaround for directed cyclic graph using Strongly Connected Component (SCC) extraction. Numbers inside nodes indicates node index. Casual edges that could be used by GCC are highlighted in green, and the opposite are orange colored. Best viewed in color. 
  }
  \label{fig:ci-example}
\end{figure}

Specifically, we can preserve these non-causal edges by performing a \emph{topology sorting} on the original graph, which makes all edges to be causal.
An example is shown in \autoref{fig:ci-topo}, we can observe that node-3 and node-4 are swapped by topology sorting, and  $\text{edge}_{4\rightarrow3}$ that should have been originally removed by GCC is preserved as $\text{edge}_{3\rightarrow4}$ after sorting. However, topology sorting requires that the graph should be an \emph{acyclic} graph. To satisfy this requirement while preserving more context information, we need to remove as few edges as possible before topology sorting, namely the maximum spanning directed acyclic graph problem. 
Such a problem has been studied in previous researches~\cite{DBLP:conf/starsem/Schluter15} and has been acknowledged as an NP-hard problem. To address this issue, we further propose a simple but efficient solution.

For scene graphs, a cycle tends to appear within a small group of objects that perform similar actions. It indicates that components with cycles, which form strongly connected components, are relatively small compared to the whole graph, as shown on the left side of \autoref{fig:ci-reorder}. 
Based on this observation, a simple workaround is applied to bypass the acyclic requirement in topology sorting. We present an example of this workaround in \autoref{fig:ci-reorder}.
Specifically, by extracting all strongly connected components (SCC) from the graph, we can construct an acyclic condensed graph according to all extracted SCC, which can be sorted in topological order. In this way, we can perform topology sorting for most of the original graph nodes and obtain a reordered graph with more causal edges. The reordered graph is then compressed by our proposed framework. By virtue of the above prepossessing, we can significantly reduce the context losses resulting from GCC and thus preserve more context information of scene graphs.  
Note that the sorting indices are not included in the compressed data, because by reordering nodes in the graph, the topology of the new graph still equals the original one.

We adopt Kosaraju's algorithm\cite{SHARIR198167} to implement the extraction of the strongly connected components. The reason for using this algorithm is that it could extract components and sort the extracted components in topological order in one step. The sorted components are then lined up to form a sorting sequence, which is used to reorder the original graph. 
The detailed process is described in \autoref{alg:scc-sorting}.

Naturally, further sorting could be performed on the extracted strongly connected components to achieve further improvement with a brute-force method, but in many cases this step is unnecessary, as most nodes have already been sorted by the proposed method, and further sorting will not bring remarkable improvement. More details can be shown in \autoref{sec:experiment-abl-ci}.

\begin{algorithm}[t]
\small
  \caption{The proposed preprocessor for context improvement.}
  \label{alg:scc-sorting}
  \LinesNumbered
  \KwIn{input graph $\mathbf{G}$ with nodes $ \mathbf{V} $ and edges $\mathbf{E}$}
  \KwOut{output sorted graph $\mathbf{G'}$ with nodes $ \mathbf{V'} $ and edges $\mathbf{E'}$}

  Initialize SortingIndices as empty list
  

  SCC = KosarajuSCC($\mathbf{G}$)

  \tcp{Build Sorting Indices}

  \ForEach{VertexSet $\mathbf{V}_{\{\mathbf{I}\}}$ in SCC}{
      Append(SortingIndices, $\{\mathbf{I}\}$)
  }


  $\mathbf{G}'$ = ReorderNodes($\mathbf{G}$, SortingIndices)

  \Return{$\mathbf{G}'$}

\end{algorithm}

\section{Learned Distribution Module}
\label{sec:method-dm}
In this section, we introduce the design of the distribution module in the element predictors. It receives the parameters generated by the parameter estimations and outputs the final probability $P(\mathbf{x})$ according to the data distribution for the lossless entropy coders.
Specifically, we consider two cases of data distribution: (1) the \emph{categorical} data, and (2) the \emph{numerical} data.
As the numerical data in scene graph may be under complicated constraints, we further propose a learned distribution model for numerical data to improve prediction.

\textbf{(1) Categorical Data.}
For \emph{categorical} data such as object types and relation types, we can utilize a direct distribution model, 
where the given distribution parameters directly represent symbol probabilities. 
In this way, the element predictor works similarly to a classification model. The entropy of such a distribution model could be represented as the cross-entropy between estimated probability and the `ground truth' value given by the symbol to compress.

\textbf{(2) Numerical Data.}
For \emph{numerical} data such as object locations, using an indirect distribution model to fit the unknown continuous distribution is a common solution. The indirect distribution model defines a cumulative distribution function constructed by the given distribution parameters, and symbol probabilities are then sampled from this function. Importantly, this distribution model is expected to fit any possible data distribution to achieve the best prediction result.
Data with simple distributions can be well fitted by Gaussian distribution. As for data with more complex distributions, we can adopt a mixture distribution composed of a series of simple distributions, such as the Gaussian mixture distribution. However, in scene graph data, numerical data such as object locations may deliver very complicated distributions. For example, for a simple scene graph where a person rides a bicycle behind a car, the person's location should be constrained to be above the bicycle and behind the car, which depicts a single-sided distribution constraint. In reality, an object in the scene graph may be correlated to multiple other objects, thus indicating more complicated distribution constraints.
In such cases, simple distributions and mixture distributions may not fit well, leading to sub-optimal probability prediction.


We propose a learned distribution model to amend the drawbacks of simple distributions and mixture distributions discussed above.
This model $c(\mathbf{x}, \mathbf{\omega})$ defines a cumulative distribution function, where $\mathbf{\omega}$ denotes the parameters of neural network.
In this way, the cumulative distribution function could be represented as:
\begin{eqnarray}
  \label{eqn:gcrd-fulldyn}
  C_{\mathbf{\omega}}(\mathbf{x}) = c(\mathbf{x}, \mathbf{\omega})
\end{eqnarray}
Nevertheless, the network parameters in $c(\mathbf{x}, \mathbf{\omega})$ are fully dynamic, which may make it hard to be optimized. 
To address this issue, we further split the network into two parts $c_1$ and $c_2$. The parameters of $c_{1}$ are initialized and optimized in the traditional way. Only the parameters of $c_{2}$ are given by the parameter estimation module to adjust the distribution. Therefore, the cumulative distribution function could then be represented as:
\begin{eqnarray}
  \label{eqn:gcrd-dyn}
  C_{\mathbf{\omega}}(\mathbf{x}) = c_{2}(c_{1}(\mathbf{x}, \mathbf{\theta}), \mathbf{\omega})
\end{eqnarray}
Compared with the fully-dynamic model, the above split can stabilize the optimization of the whole model. In practice, the univariate non-parametric density model proposed by \cite{DBLP:conf/iclr/BalleMSHJ18} is utilized as the implementation for $c(\mathbf{x}, \mathbf{\omega})$. 

\section{Experiments}
\label{sec:experiments}
In this section, we conduct extensive experiments to validate the effectiveness of our proposed framework.

\subsection{Experimental Settings}
\subsubsection{Datasets}
\label{sec:datasets}
In order to show the proposed framework works with different kinds of scene graphs, four datasets are selected. Those datasets are created in different ways, thus have different partitions of data elements and diverse density of structure.
\begin{itemize}
  \item \textbf{VisualGenome (VG)\cite{DBLP:journals/ijcv/KrishnaZGJHKCKL17}}: A widely used dataset in scene graph generation works, including 10K images, each with annotated scene graphs. Following \cite{DBLP:conf/cvpr/XuZCF17}, the most frequent 150 object types and 50 relation types are selected, while all other objects and relations are discarded.
  \item \textbf{VisualGenome generated by scene graph generator (VG-Gen)}: Based on VisualGenome images, we apply existing scene graph generation methods to generate new scene graph datasets for experiment. This resembles real-life application of the proposed method. Scene Graph Benchmark\cite{han2021image}, an open-source software including many scene graph generation methods, is use to generate the dataset.
  \item \textbf{VisualGenome generated by relation prediction (VG-GenRel)}: Different from VG-Gen, we set the relation head in Scene Graph Benchmark\cite{han2021image} to `\textit{predcls}' mode to generate VG-GenRel. This mode reuses existing annotated labels of objects in VG as object detection results, and just generates relation data.
  \item \textbf{HiEve\cite{DBLP:journals/corr/abs-2005-04490}}: A dataset containing annotated human objects with action labels under complex events. Scene graphs are mostly generated with (and some are manually labeled from) those annotated actions, by densely connect all objects that perform the same group-related actions with undirected edges. Objects that do not belong to such actions are removed.
\end{itemize}

The statistics of all datasets is shown in \autoref{tab:dataset-stats}. 
We also present samples of each dataset in \autoref{fig:dataset-showcase}.
It can be seen that the VG dataset, which is created by manually labeling objects and relations, contains mainly node data. The generated VG-Gen has larger graphs with balanced node and edge data, and VG-GenRel is edge data-centric with denser graphs. HiEve is much smaller than VG, but its graphs are very dense.

\begin{table}[!t]
  \centering
  \caption{Statistics of used datasets.}
  \label{tab:dataset-stats}
  \begin{minipage}{1.\linewidth}
    \centering
    \setlength{\tabcolsep}{6pt}
    \begin{tabular}{lllll}
      \hline
      Dataset & Nodes & Edges & Graphs & Density\footnote{Average graph density, calculated for each graph by $\dfrac{edges}{nodes^{2}} $} \\
      \hline
      VG & 1.63M & 0.89M & 95K & 5.48\% \\
      VG-Gen & 8.70M & 10.8M & 109K & 2.55\% \\
      VG-GenRel & 1.04M & 4.14M & 88K & 34.58\% \\
      HiEve & 9.7K & 81.8K & 1.6K & 71.10\% \\
      \hline
      \end{tabular}
  \end{minipage}
\end{table}

\begin{figure}[!t]
  \centering
  \setlength{\tabcolsep}{0pt}
  \small
  \begin{tabular}{ccc}
    \vspace{-10pt}
    (a) & 
    \begin{minipage}[]{.24\textwidth}
      \centering
      \includegraphics[width=1\textwidth, height=0.75\textwidth]{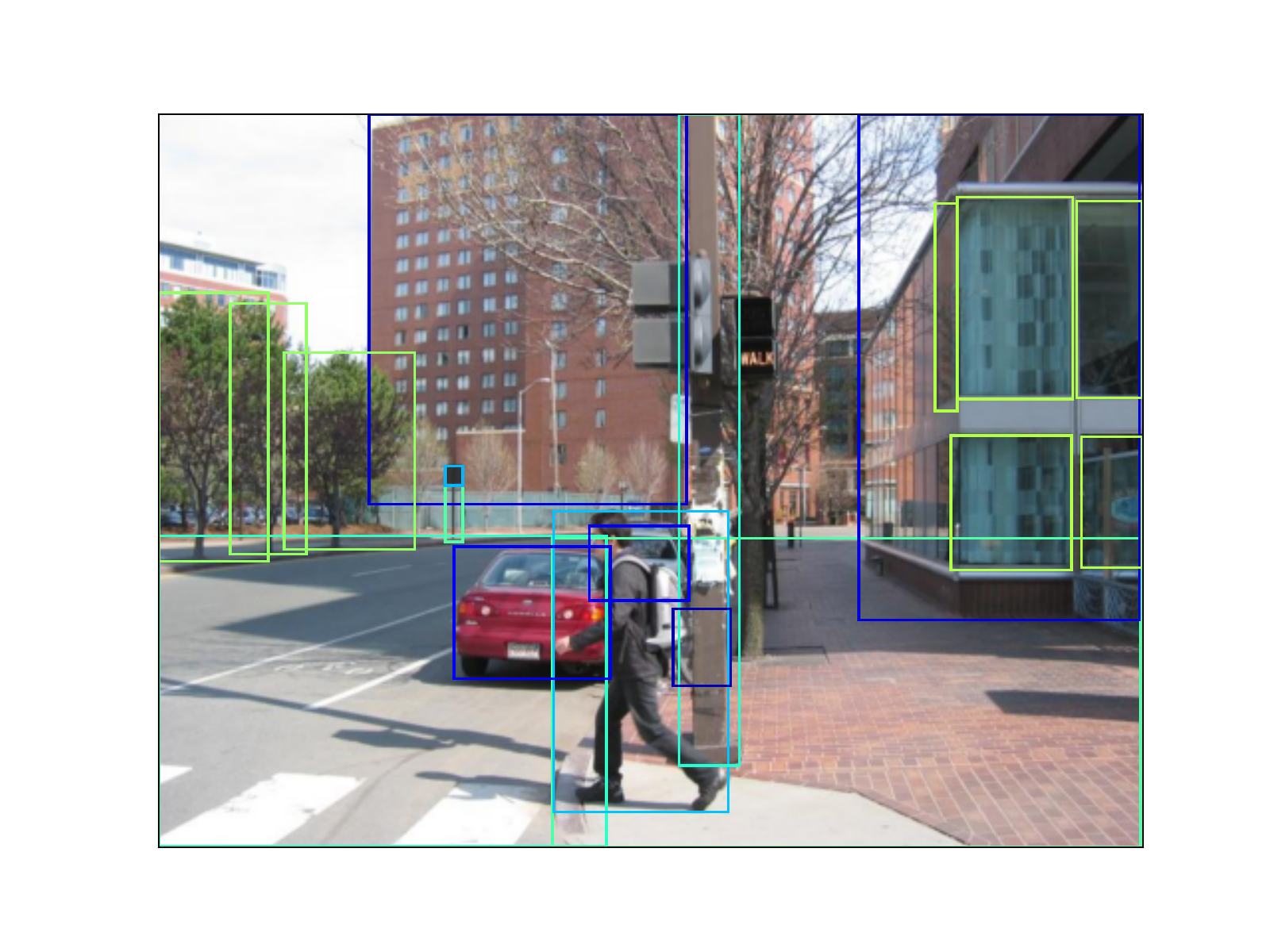}
    \end{minipage}
    & 
    \begin{minipage}{.24\textwidth}
      \includegraphics[width=1\textwidth]{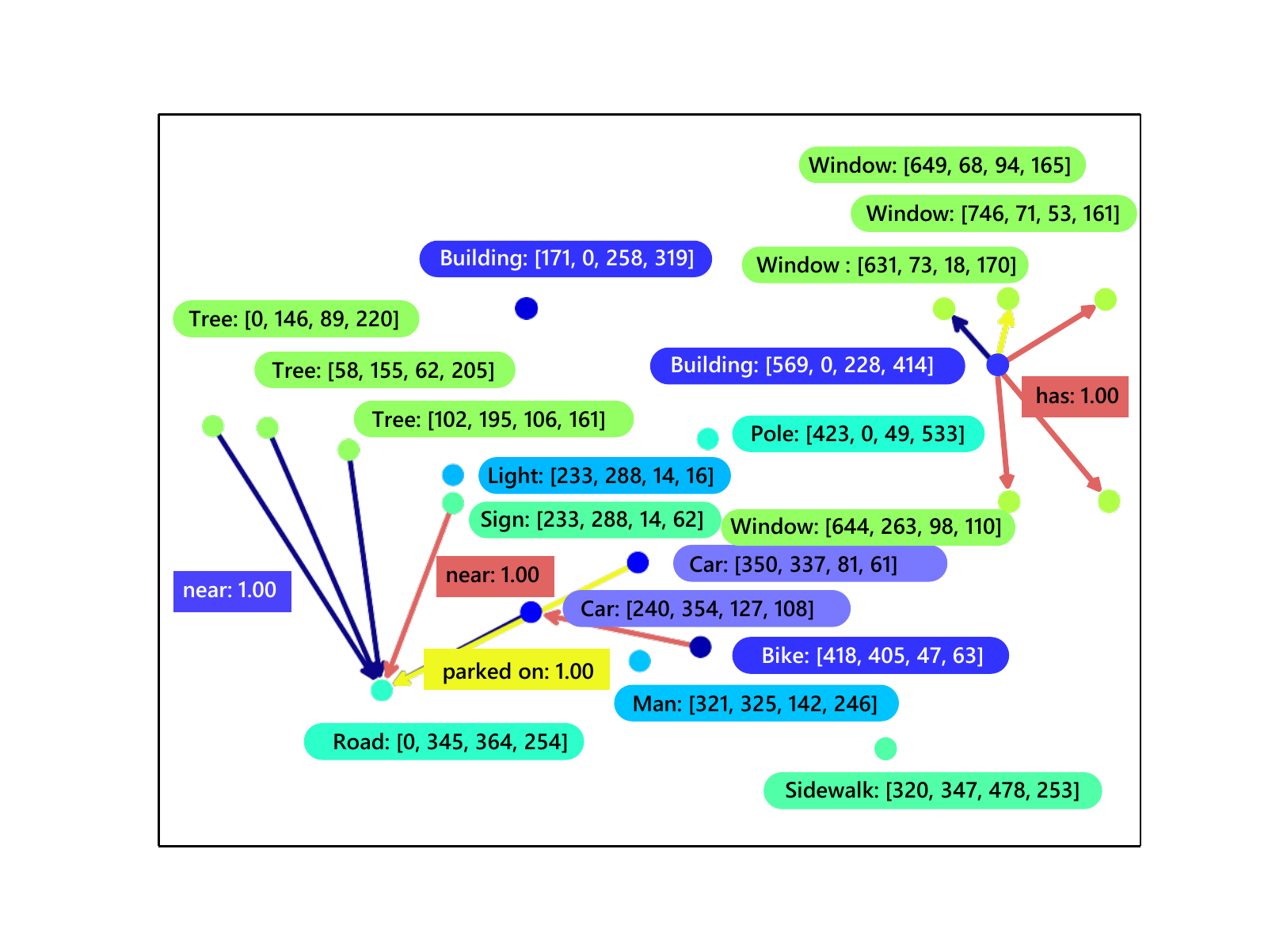} 
    \end{minipage} \\
    \vspace{-10pt}
    (b) & 
    \begin{minipage}{.24\textwidth}
      \centering
      \includegraphics[width=1\textwidth, height=0.75\textwidth]{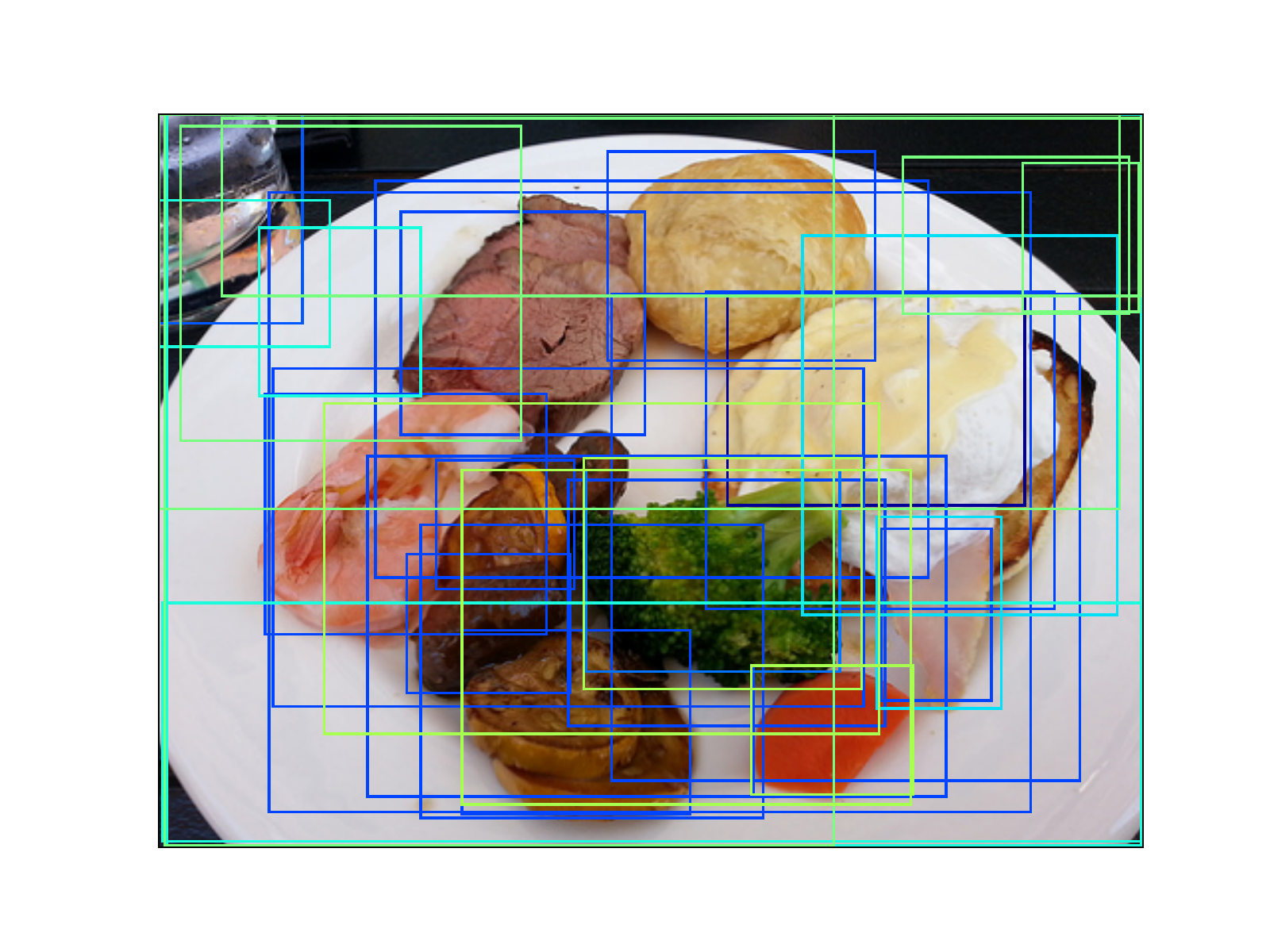}
    \end{minipage} & 
    \begin{minipage}{.24\textwidth}
      \includegraphics[width=1\textwidth]{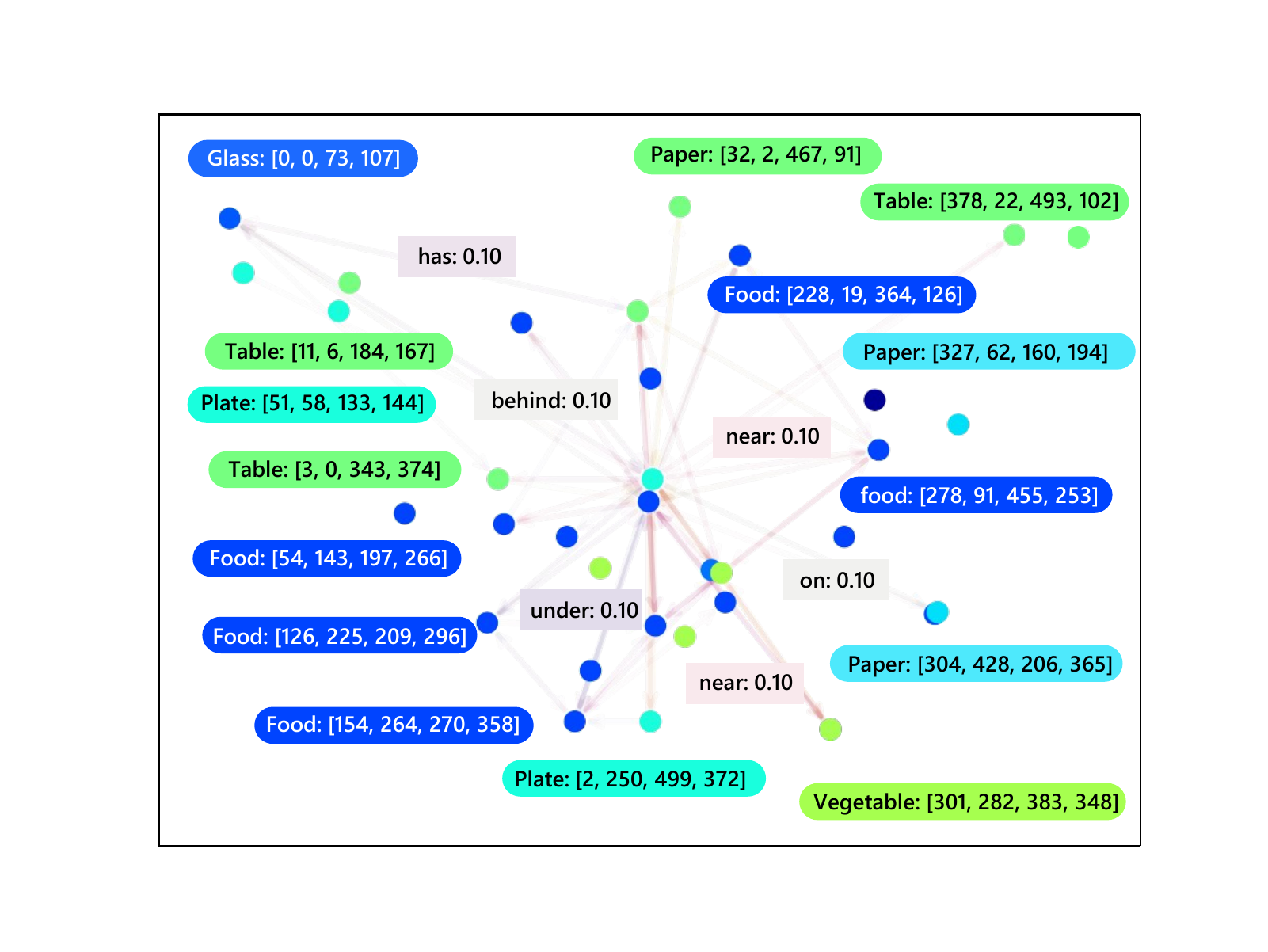} 
    \end{minipage} \\
    \vspace{-10pt}
    (c) & 
    \begin{minipage}{.24\textwidth}
      \centering
      \includegraphics[width=1\textwidth, height=0.75\textwidth]{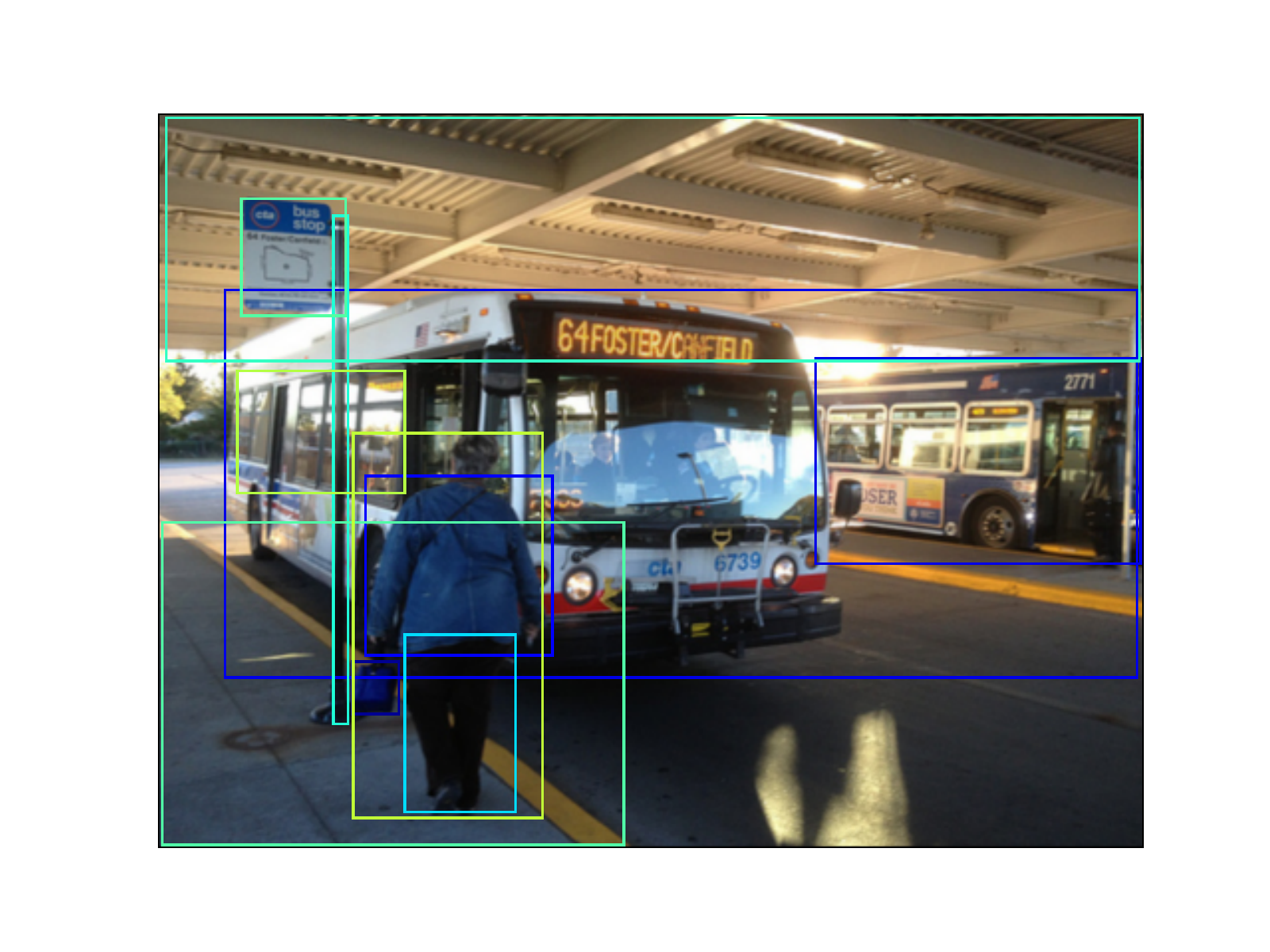}
    \end{minipage} & 
    \begin{minipage}{.24\textwidth}
      \includegraphics[width=1\textwidth]{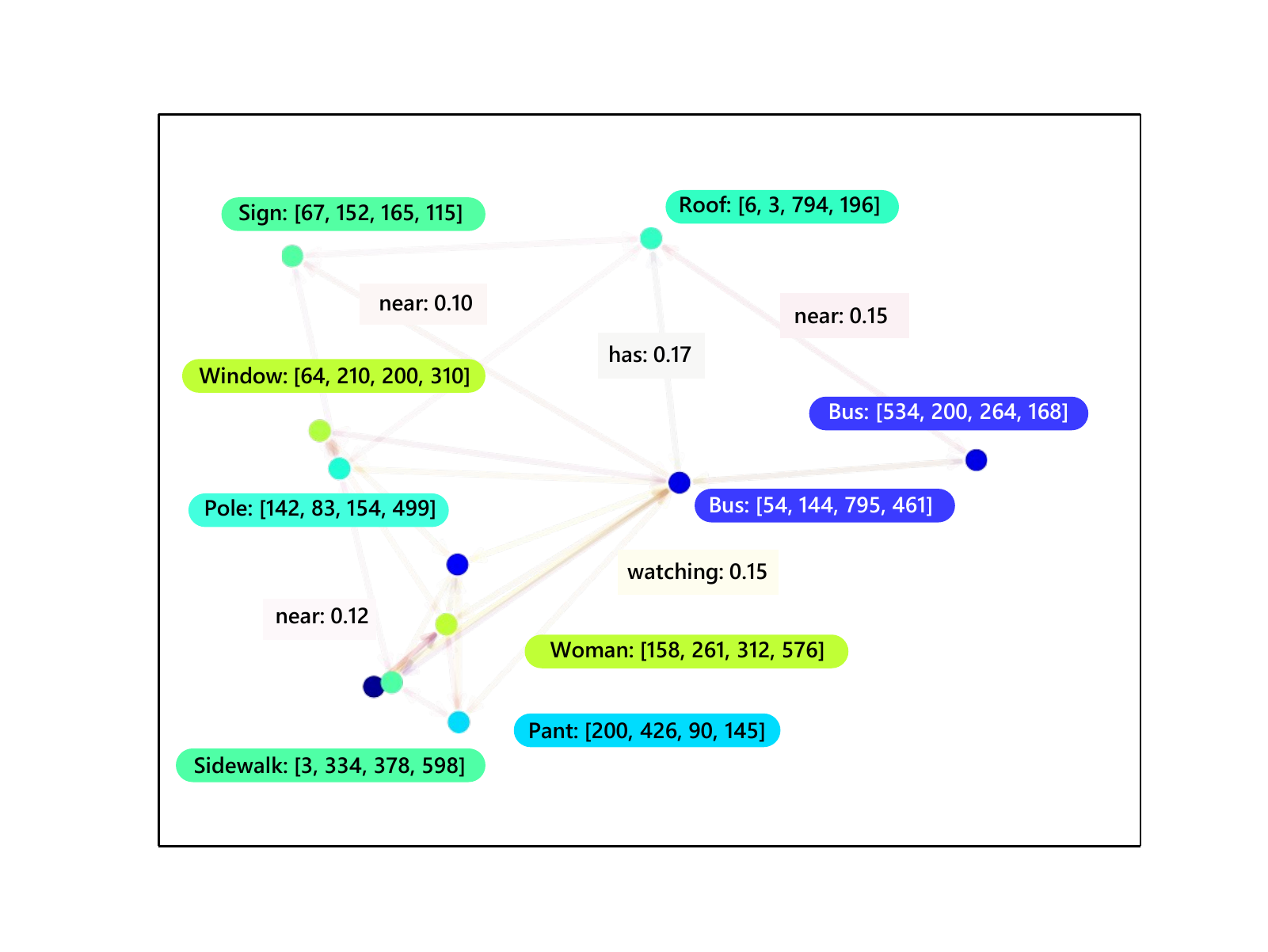} 
    \end{minipage} \\
    \vspace{-10pt}
    (d) & 
    \begin{minipage}{.24\textwidth}
      \centering
      \includegraphics[width=1\textwidth, height=0.75\textwidth]{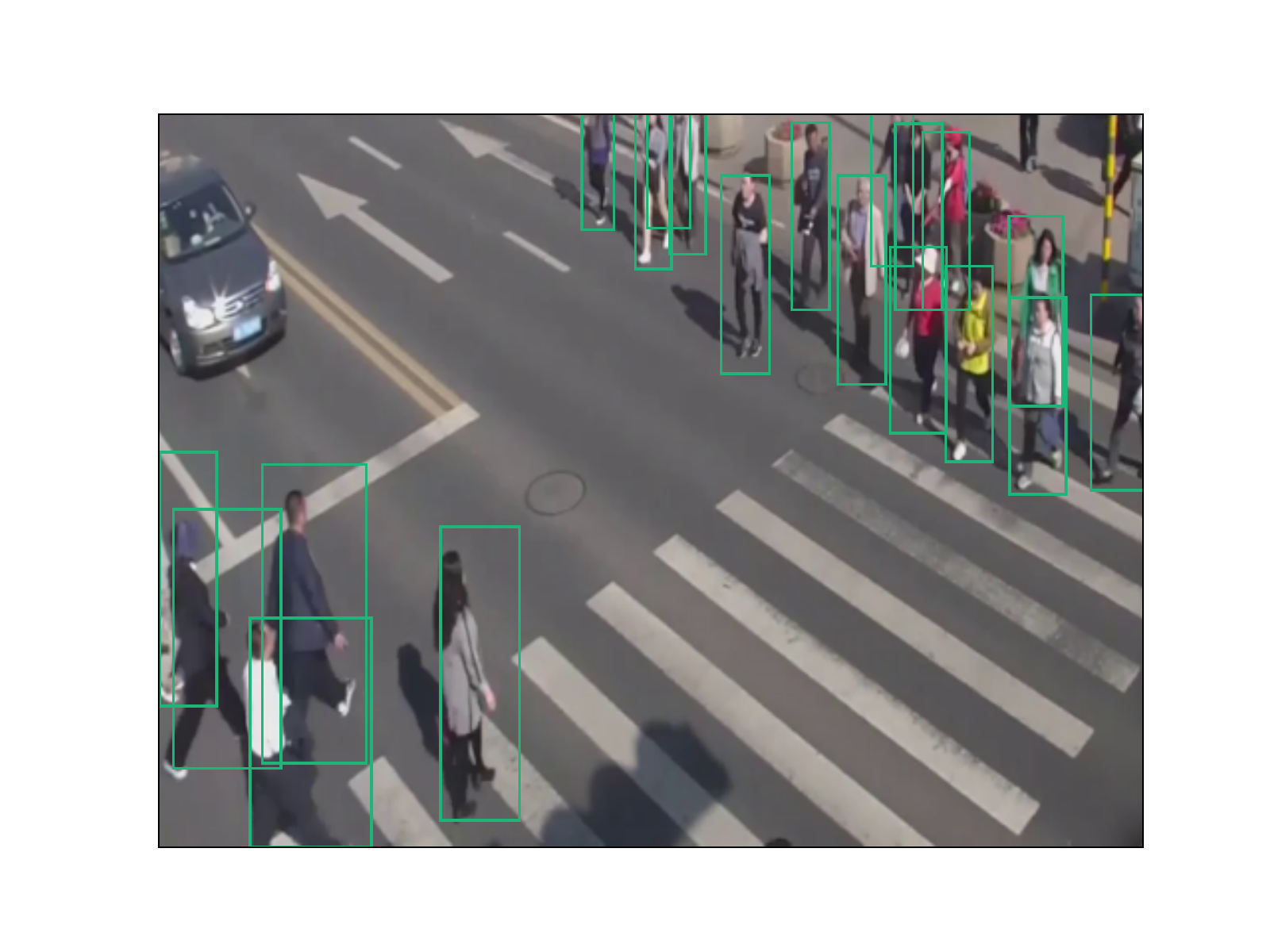}
    \end{minipage} & 
    \begin{minipage}{.24\textwidth}
      \includegraphics[width=1\textwidth]{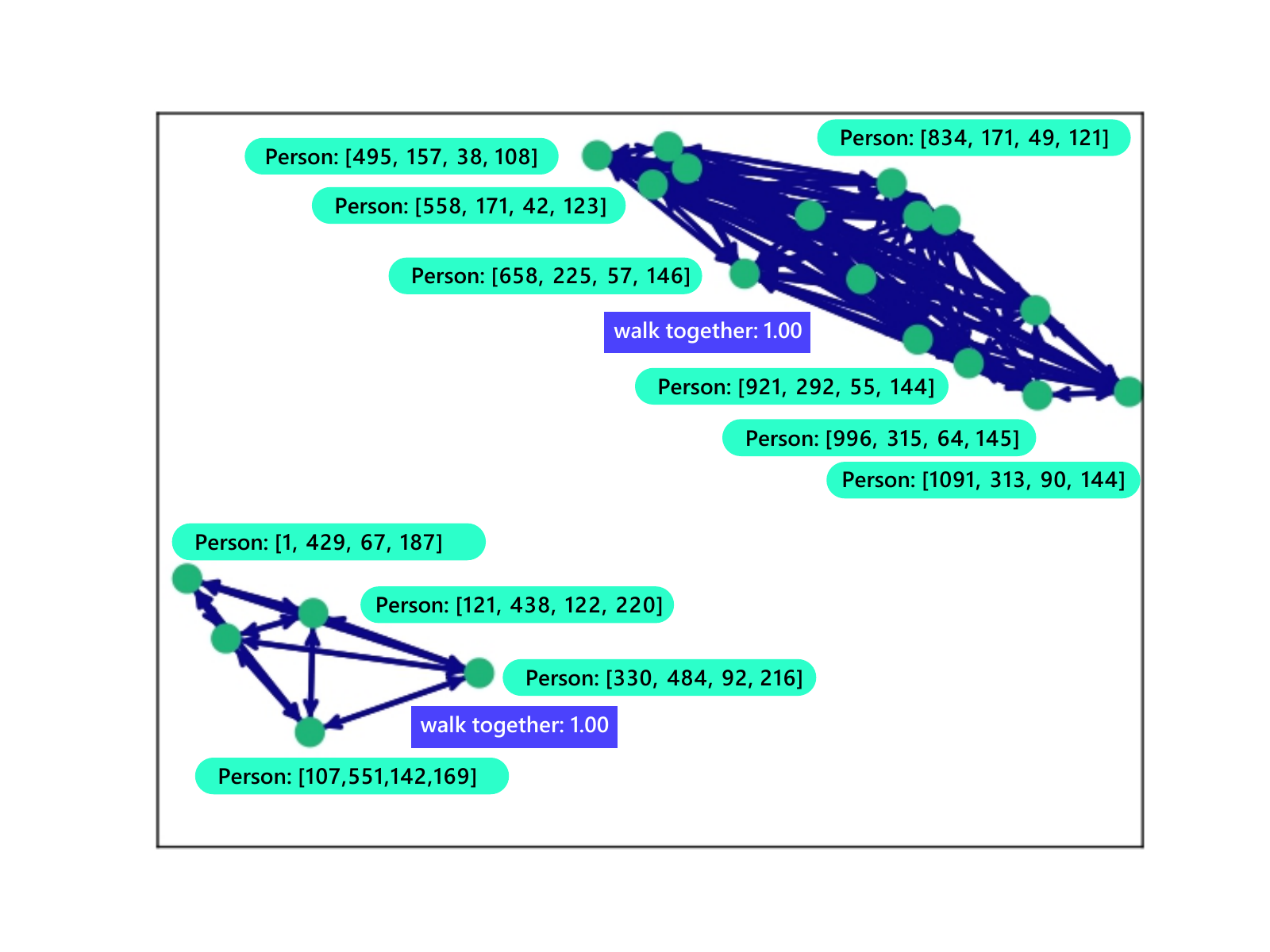} 
    \end{minipage} \\
  \end{tabular}

  \caption{Visualization of a sample image (left) and scene graph (right) in each dataset: (a) VG (b) VG-Gen (c) VG-GenRel (d) HiEve (objects not performing group-related actions are ignored). 
  For scene graph samples, the position of the nodes indicate the bounding box center of the objects. 
  Different colors of nodes and edges indicate objects and relations with different types. Note that in (d) HiEve the grey nodes indicate objects not included in the scene graph due to not performing group-related actions. 
  The transparency of the edges indicate relation weights. 
  Due to too much data in a single image, only a few data labels in those graphs are displayed.
  }
  \label{fig:dataset-showcase}
\end{figure}

\subsubsection{Metrics}
\begin{itemize}
    \item \textbf{Compression ratio}. We use the compression ratio as our main metric for lossless scene graph compression. It is calculated by dividing the length of the compressed bitstream by the length of the original bitstream. A lower ratio indicates a better compression.
    
    \item \textbf{Bits Per Node}. As scene graph compression is closely related to the graph compression task, we also adopt the `Bits Per Node' metric, which is usually applied in graph compression works\cite{DBLP:conf/www/BoldiV04,DBLP:conf/spire/BrisaboaLN09}.
\end{itemize}

\subsubsection{Implementation Details}
The unified prior extractor is implemented with an autoencoder with both the encoder and decoder composed of 4 fully-connected layers as well as ReLU activation layers, and the latent vector channels setting to 2. 
For the element predictors,
the GCC-based models are composed of 1 GCC layer followed by 3 fully-connected layers as well as ReLU activation layers. 
For the entropy coders, Asymmetric Numeral System\cite{DBLP:journals/corr/Duda13} is adopted as the implementation.
Moreover, other meta-element in datasets that has not been compressed in the experiments, such as image sizes, are simply compressed with the zlib\cite{zlib} compressor.

For all datasets, 80\% of the scene graphs are used as the training set and 20\% as testing set. Note that for HiEve dataset, as the generated scene graph is undirected, which means that the adjacency matrix of the graph structure is symmetric, only the upper-triangle part of the adjacency matrix as well as the corresponding relation types are compressed. 
The experiments are conducted with Intel i7 4700k CPU and 12GB NVIDIA GTX TITAN Xp GPU.

\subsubsection{Competitors}
We compare our proposed framwork with two streams of existed approaches:
\begin{itemize}
    \item \textbf{General-purpose compression methods}. We first choose some widely used general-purpose codecs for comparison, including zlib\cite{zlib}, LZMA\cite{LZMA}, Brotli\cite{DBLP:journals/tois/AlakuijalaFFKOS19} and LPAQ1\cite{Mahoney2005AdaptiveWO}. Moreover, a combined framework (called `Combined') with the mixture of traditional specialized codecs for different elements of scene graph is compared. The combined framework use WebGraph\cite{DBLP:conf/www/BoldiV04} method for compressing graph structure, and Zstd\cite{Zstd} for node data and edge data. Besides, Dzip\cite{DBLP:conf/dcc/GoyalTCO20}, one of the state-of-the-art learning-based general-purpose compression framework, is used for comparison.
    \item \textbf{Graph structure compression methods}. Previous graph lossless compression methods like WebGraph\cite{DBLP:conf/www/BoldiV04} and k2-tree\cite{DBLP:conf/spire/BrisaboaLN09} cannot be directly applied to scene graph compression because of extra node and edge data. However, as the graph structure is the first graph element to decompress in the proposed method, the predictor and entropy coder for graph structure could be substituted by other graph structure lossless compression methods. In this way, we can compare our proposed method with a predictor based on the proposed Directed Graph Context Autoencoder (DGCAE) and other graph structure lossless compression methods in graph structure compression. When reporting the performance for graph structure compression, the random node sorting is applied to simulate different object orders in scene graphs.
\end{itemize}

\subsection{Comparison Results}
\label{sec:experiments-results}
In this section, according to the type of competitors, we report the overall performances from two aspects.
\subsubsection{Whole Scene Graph Compression} The compression ratio of our proposed framework and related competitors for the whole scene graphs are summarized in \autoref{tab:comp-total}. It's obvious that our proposed framework outperforms all traditional general-purpose compressors with a significant margin on all benchmarks. The main reason for this is that our proposed frameworks is specially designed for whole scene graph compression using element-adaptive predictors. It can also efficiently exploit graph structure to reduce context redundancy. 

\begin{table*}
  \centering
  \scriptsize
  \setlength{\tabcolsep}{4pt}
  \caption{Compression Ratio for total scene graph compression of different compression frameworks in different datasets.}
  \label{tab:comp-total}
    \begin{tabular}{c|cc|cccc}
    \hline
    Methods & Graph-based & Training & VG & VG-Gen & VG-GenRel & HiEve \\
    \hline
    zlib\cite{zlib} &  &  & 70.58\% & 44.64\% & 46.02\% & 22.79\% \\
    LZMA\cite{LZMA} &  &  & 129.25\% (83.14\%) & 52.38\% (17.34\%) & 80.24\% (74.33\%) & 43.97\% (92.97\%) \\
    Brotli\cite{DBLP:journals/tois/AlakuijalaFFKOS19} &  &  & 64.28\% (-8.92\%) & 38.92\% (-12.81\%) & 43.31\% (-5.90\%) & 19.49\% (-14.47\%) \\
    LPAQ1\cite{Mahoney2005AdaptiveWO} &  &  & 58.01\% (-17.80\%) & 30.00\% (-32.79\%) & 40.59\% (-11.81\%) & 18.22\% (-20.05\%) \\
    Combined (\cite{Zstd}+\cite{DBLP:conf/www/BoldiV04}) & $\surd$ &  & 65.84\% (-6.71\%) & 39.55\% (-11.41\%) & 39.37\% (-14.47\%) & 18.94\% (-16.87\%) \\
    Dzip\cite{DBLP:conf/dcc/GoyalTCO20} &  & $\surd$ & 44.21\% (-37.35\%) & 34.41\% (-22.93\%) & 25.11\% (-45.44\%) & 13.20\% (-42.06\%) \\
    Ours & $\surd$ & $\surd$ & \textbf{30.19\% (-57.23\%)} & \textbf{26.84\% (-39.86\%)} & \textbf{16.99\% (-63.08\%)} & \textbf{8.52\% (-62.63\%)} \\
    \hline
    \end{tabular}
\end{table*}

\subsubsection{Graph Structure Compression}
\label{sec:abl-gc}
The bits per node metric produced by different compression methods are summarized in \autoref{tab:abl-gc}. It can be observed that our proposed method surpasses all the traditional methods in graph structure compression. Meanwhile, we notice that the compression ratio of the proposed method is close to traditional graph compression methods on some datasets like VG and VG-Gen. This is because our proposed method encodes the adjacency matrix in a per-symbol manner, which may not be the most efficient way to compress sparse matrices. Unfortunately, most scene graphs in those datasets are relatively sparse, according to \autoref{tab:dataset-stats}. 
This could be further improved in future works. Nevertheless, for other datasets with dense graphs, this drawback is largely resolved, and the superiority of the proposed framework is therefore prominent.

\begin{table}
  \centering
  \scriptsize
  \caption{Bits Per Node metric of the relation link data bitstream (and relative reduction) produced by different graph structure compression methods in different datasets.}
  \label{tab:abl-gc}
  \begin{tabular}{c|cc}
    \hline
    \diagbox{Methods}{Datasets} & VG & VG-Gen \\
    \hline
    WebGraph\cite{DBLP:conf/www/BoldiV04} & 0.62 & 1.24 \\
    k2tree\cite{DBLP:conf/spire/BrisaboaLN09} & 0.60 (-3.40\%) & 1.54 (23.61\%) \\
    Ours & \textbf{0.54 (-13.10\%)} & \textbf{1.09 (-12.17\%)} \\
    \hline
    \diagbox{Methods}{Datasets} & VG-GenRel & HiEve \\
    \hline
    WebGraph\cite{DBLP:conf/www/BoldiV04} & 2.86 & 4.46 \\
    k2tree\cite{DBLP:conf/spire/BrisaboaLN09} & 2.16 (-24.49\%) & 3.45 (-22.79\%) \\
    Ours & \textbf{1.69 (-40.72\%)} & \textbf{1.40 (-68.62\%)} \\
    \hline
    \end{tabular}
\end{table}

\subsection{Ablation Study on the Entire Framework}
In order to study the effectiveness of different modules in our proposed framework, we conduct experiments by removing some modules or replacing some of them with other solutions to gain further insight. First, the proposed joint compression framework could be replaced by other learning-based general-purpose compression frameworks such as Dzip\cite{DBLP:conf/dcc/GoyalTCO20}. 
Second, the graph context modeling techniques as well as the corresponding preprocessor for context improvement discussed in \autoref{sec:method-cm} can be removed so that each element predictor uses only the initial prediction result from the unified prior extractor. Finally, the learned distribution model discussed in \autoref{sec:method-dm} could be replaced by a simple Gaussian distribution model. Experiments are then conducted with the above variants. Results are shown in \autoref{tab:abl-fw}.

\begin{table}
  \centering
  \scriptsize
  \setlength{\tabcolsep}{4pt}
  \caption{Results of removing or replacing different modules in the proposed framework.}
  \begin{minipage}{1.\linewidth}
    \centering
    \begin{tabular}{c|c|c}
    \hline
    Datasets & Framework & \thead{Compression \\ Ratio (Relative)} \\
    \hline
    \multirow{4}{*}{VG} & Dzip\cite{DBLP:conf/dcc/GoyalTCO20} & 44.21\% \\
     & Ours - Context - LearnDist & 33.50\% (-24.24\%) \\
     & Ours + Context - LearnDist & 30.93\% (-30.05\%) \\
     & Ours + Context + LearnDist & \textbf{30.19\% (-31.73\%)} \\
     \hline
    \multirow{4}{*}{VG-Gen} & Dzip\cite{DBLP:conf/dcc/GoyalTCO20} & 34.41\% \\
     & Ours - Context - LearnDist & 30.43\% (-11.56\%) \\
     & Ours + Context - LearnDist & 27.24\% (-20.84\%) \\
     & Ours + Context + LearnDist & \textbf{26.84\% (-21.98\%)} \\
     \hline
    \multirow{4}{*}{VG-GenRel} & Dzip\cite{DBLP:conf/dcc/GoyalTCO20} & 25.11\% \\
     & Ours - Context - LearnDist & 20.80\% (-17.17\%) \\
     & Ours + Context - LearnDist & 17.27\% (-31.21\%) \\
     & Ours + Context + LearnDist & \textbf{16.99\% (-32.34\%)} \\
     \hline
    \multirow{4}{*}{HiEve} & Dzip\cite{DBLP:conf/dcc/GoyalTCO20} & 13.20\% \\
     & Ours - Context - LearnDist & 12.24\% (-7.27\%) \\
     & Ours + Context - LearnDist & 8.56\% (-35.17\%) \\
     & Ours + Context + LearnDist & \textbf{8.52\% (-35.50\%)} \\
     \hline
    \end{tabular}
\end{minipage}
\label{tab:abl-fw}
\end{table}



We can see that the proposed joint compression framework outperforms the general-purpose compression framework Dzip\cite{DBLP:conf/dcc/GoyalTCO20} in all cases, even without the graph context model and learned distribution model. This is achieved by extracting vital information from the scene graph data with the unified prior extractor and designing adaptive element predictors according to different data elements. Dzip\cite{DBLP:conf/dcc/GoyalTCO20} though, is designed for general-purpose data compression with a serial context model that does not fit graph data well and ignoring different data distributions, thus does not perform well on scene graph compression.

The graph context models also contribute significantly to the proposed framework in all datasets. The main reason is that our proposed context models are designed for all element predictors to reduce context redundancy for all data elements. 
Especially, the relative improvement is more distinct in VG-Gen, VG-GenRel, and HiEve, both of them contain generated scene graphs. It indicates that the adopted scene graph generators are more context-aware. Therefore, those generated scene graphs could benefit more from the proposed graph context models.

The proposed learned distribution model contributes less than the context model, as this module could only benefit object location compression in those datasets. Therefore, for datasets that contain more node data (\emph{e.g.}, VG and VG-Gen), the learned distribution model brings larger improvements than datasets that own fewer node data (\emph{e.g.}, VG-GenRel and HiEve).

\subsection{Ablation Study on Each Module}
In order to study the effectiveness of each key module (graph context model
, preprocessor 
and distribution model) in our proposed framework, a series of ablation studies are conducted.

\subsubsection{Graph Context Modeling}
In this experiment, the graph context model is disabled for some particular element predictors to validate its effectiveness. In order to keep along with \autoref{tab:abl-fw}, all experiments in this section are performed without the learned distribution model.
The results are summarized in \autoref{tab:abl-cm}.
Note that all experiments enables the preprocessor, including the one with all context models disabled. Thus, the result with all context models disabled in this experiment is slightly different from the "Ours - Context - LearnDist" result in \autoref{tab:abl-fw}.

\begin{figure}[!t]
  \centering
   \subfloat[]{
    \includegraphics[width=0.24\textwidth]{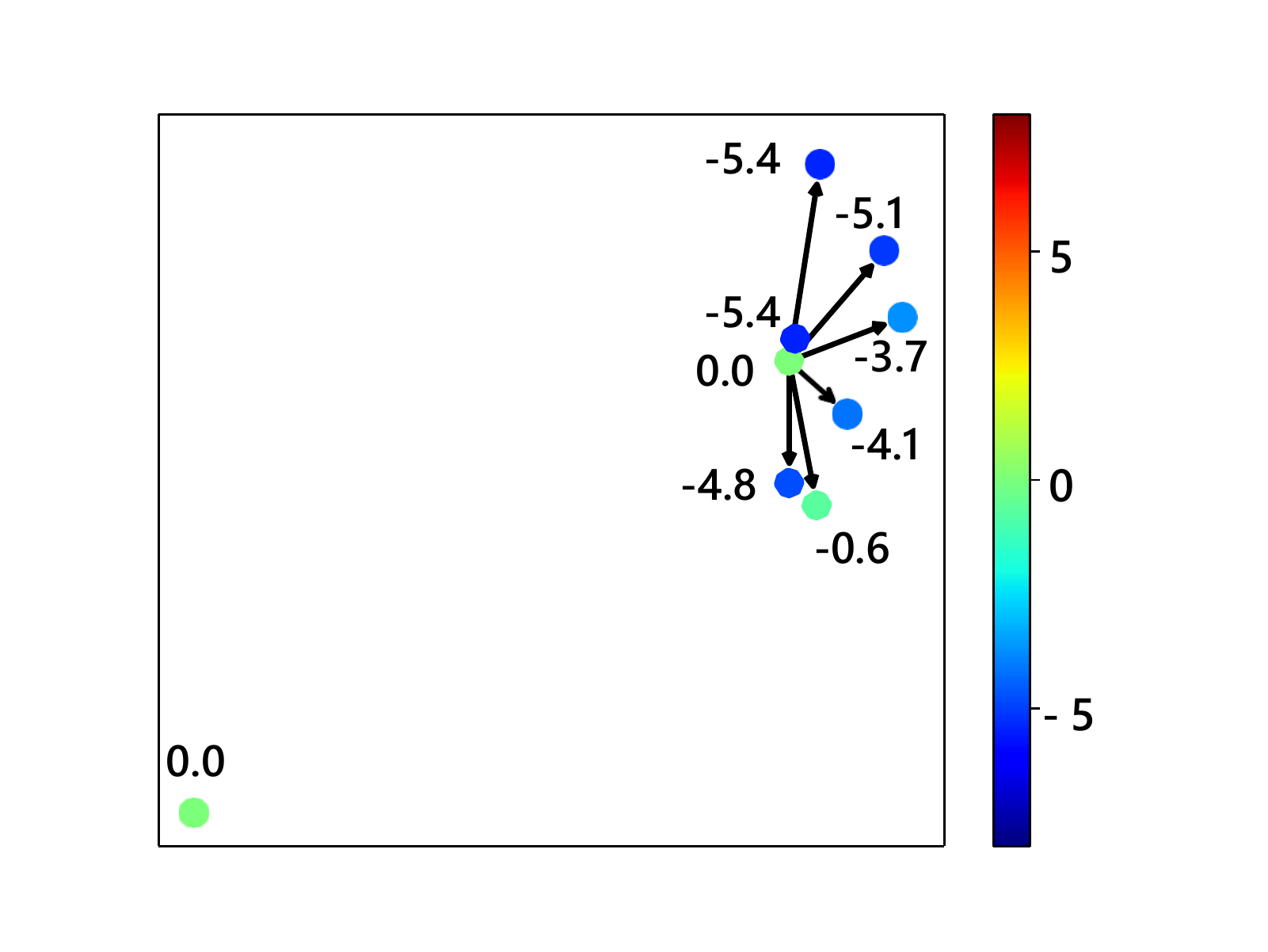}
     \label{fig:exp-context:a}
    }
   \subfloat[]{
    \includegraphics[width=0.24\textwidth]{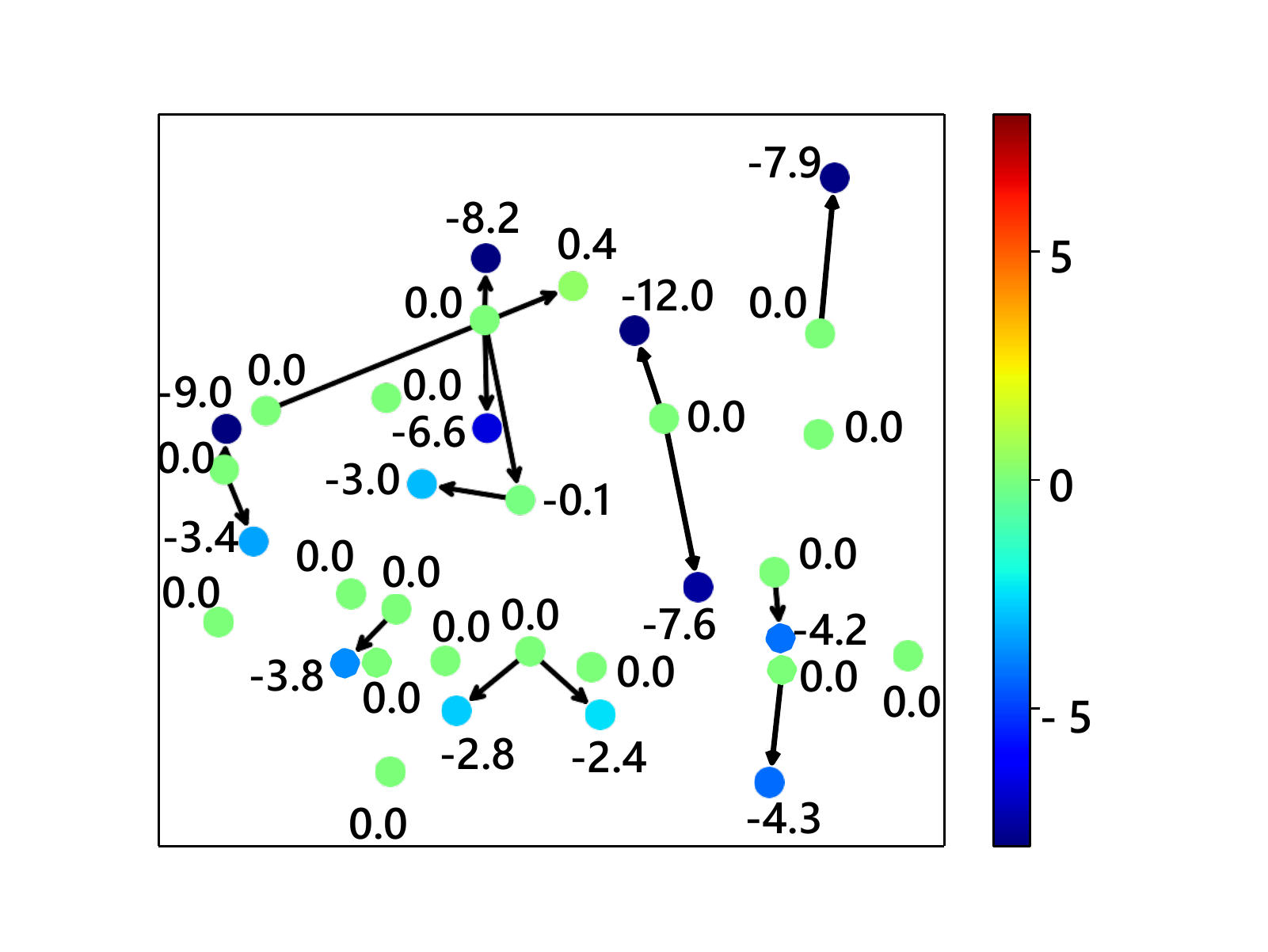}
     \label{fig:exp-context:b}
   }
  \caption{Qualitative results to show improvement by the node context model from the object location predictor. The node color and the labeled numbers in those figures indicates the entropy decrease for this node brought from context information. Therefore, blue color indicate that the compression of this node is improved with context information, while red color indicates the opposite. Green color usually indicate zero improvement because of no context information to utilize.
  }
  \label{fig:exp-context}
\end{figure}

\begin{table}
  \centering
  \scriptsize
  \setlength{\tabcolsep}{4pt}
  \caption{Results of disabling context models in particular element predictors.}
  \begin{minipage}{1.\linewidth}
    \centering
    \begin{tabular}{c|ccc|c}
    \hline
    Datasets  & \thead{Node\\Data} & \thead{Structure} & \thead{Edge\\Data} & \thead{Compression\\Ratio (Relative)}\\
    \hline
\multirow{4}{*}{VG} &  &  &  & 33.47\% \\
 & $\surd$ &  &  & 32.70\% (-2.32\%) \\
 & $\surd$ & $\surd$ &  & 32.39\% (-3.24\%) \\
 & $\surd$ & $\surd$ & $\surd$ & \textbf{30.93\% (-7.60\%)} \\
 \hline
\multirow{4}{*}{VG-Gen} &  &  &  & 30.48\% \\
 & $\surd$ &  &  & 27.93\% (-8.35\%) \\
 & $\surd$ & $\surd$ &  & 27.85\% (-8.64\%) \\
 & $\surd$ & $\surd$ & $\surd$ & \textbf{27.24\% (-10.63\%)} \\
 \hline
\multirow{4}{*}{VG-GenRel} &  &  &  & 20.78\% \\
 & $\surd$ &  &  & 18.48\% (-11.07\%) \\
 & $\surd$ & $\surd$ &  & 18.42\% (-11.34\%) \\
 & $\surd$ & $\surd$ & $\surd$ & \textbf{17.27\% (-16.88\%)} \\
 \hline
\multirow{4}{*}{HiEve} &  &  &  & 12.24\% \\
 & $\surd$ &  &  & 10.70\% (-12.59\%) \\
 & $\surd$ & $\surd$ &  & 10.26\% (-16.22\%) \\
 & $\surd$ & $\surd$ & $\surd$ & \textbf{8.56\% (-30.08\%)} \\     \hline
  \end{tabular}
\end{minipage}
\label{tab:abl-cm}
\end{table}
We can see that each context model contributes to the compression process in most cases. Meanwhile, the improvements from the context model in each graph element vary with datasets. This may be because different extent of contextual relationship lies beneath the data. 
For example, the structure context model brings significant improvement for HiEve, while only slight improvements on VG, VG-Gen, and VG-GenRel. 
It proves that densely-connected graphs in HiEve are more context redundant. Besides, the edge context model in HiEve also brings a great improvement. The reason may be that most humans perform the same action in a single scene in HiEve, which leads to heavy context redundancy in edge data.
Furthermore, to explicitly present the improvement brought by the context model, a \emph{qualitative experiment} is further conducted to show the disparity of probability estimation results with and without context information. 
Some results are illustrated in \autoref{fig:exp-context}. 
\autoref{fig:exp-context:a} presents a simple graph with only 7 edges. The nodes that are pointed by edges could utilize context information from the other node, thus obtaining better compression results.
\autoref{fig:exp-context:b} shows a graph with a more complicated structure.
It can be seen that although some of the nodes exploiting context information appear to have worse prediction results (with labels above zero), most nodes obtain better results. The above observation further proves the quantitative results in \autoref{tab:abl-cm}.

\subsubsection{Preprocessing}
\label{sec:experiment-abl-ci}
We apply different preprocessing methods to the graph and compare them with our proposed preprocessor based on strongly connected component extraction. 
To better evaluate the improvement of context information for different preprocessing methods, we devise the \emph{Context Improvement Rate (CIR)} metric. It compares the number of causal edges that satisfy the context condition before and after preprocessing. Assuming $\mathbf{A}$ and $\mathbf{A'}$ are the adjacency matrix before, and after preprocessing, the CIR is defined as:
\begin{eqnarray}
  CIR = \frac{\sum (\mathbf{A'} - tril(\mathbf{A'}))}{max(\sum (\mathbf{A}-tril(\mathbf{A})), 1)}
\end{eqnarray}

\begin{figure}
  \includegraphics[width=1\linewidth]{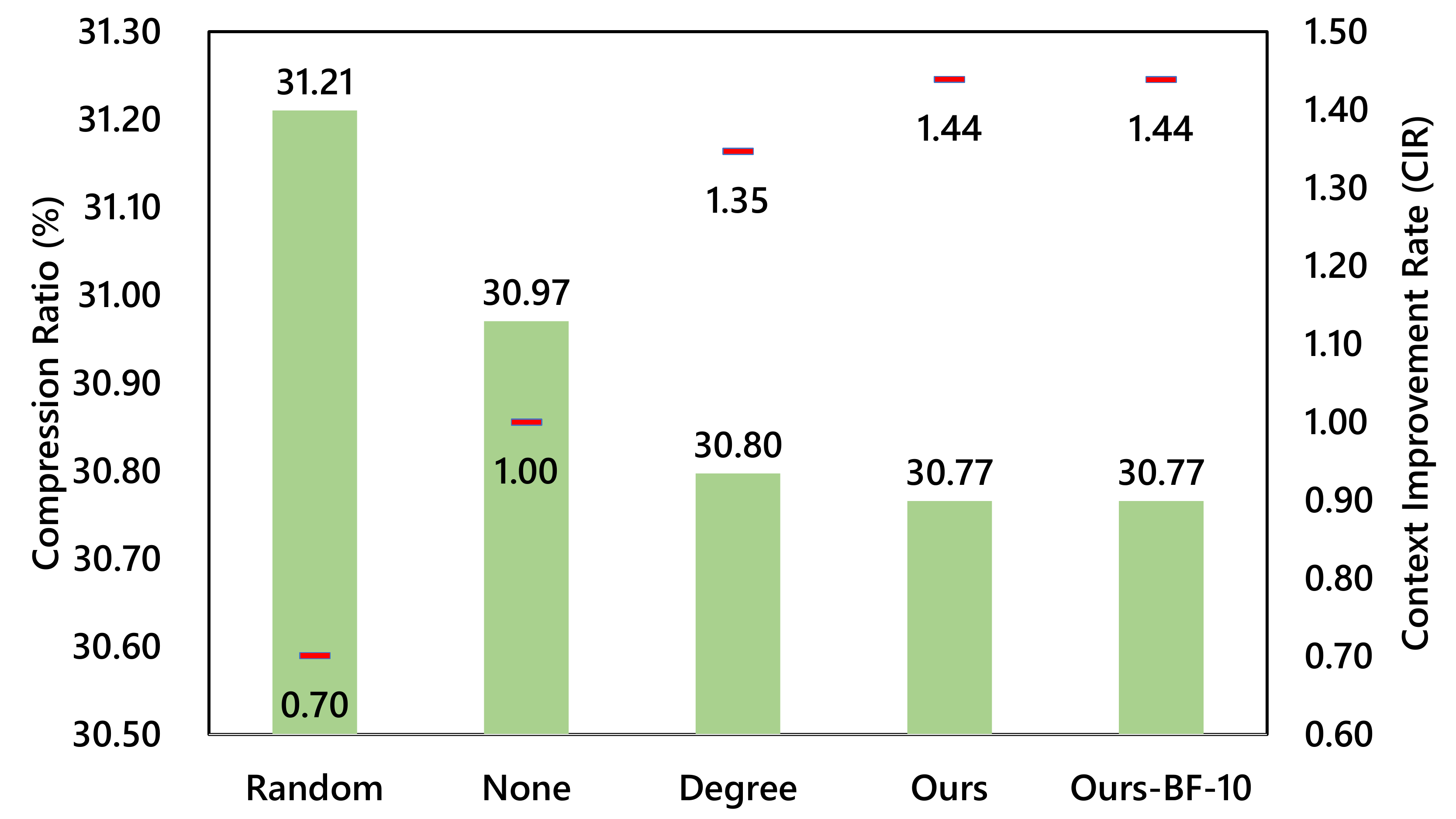}
  \caption{Results for different preprocessing methods on the VG dataset. The methods for comparison includes random sorting (Random), no sorting (None), degree-based sorting (Degree), the proposed strongly connected component with topological sorting method (Ours), and the proposed method with brute-force solver for every extracted components (Ours-BF-10). 
  }
  \label{fig:abl-ci}
\end{figure}

As the complexity of brute-force solver is $O(n!)$, the maximum size of a processed component is limited to 10 to make sure preprocessing can be completed with reasonable time and memory. And thus, all components that are larger than 10 are left unsorted. For easier comparison, we train a model checkpoint without the preprocessing step, and then all preprocessing methods are applied to this checkpoint during testing. The results are shown in \autoref{fig:abl-ci}.

\begin{figure}
  \includegraphics[width=1\linewidth]{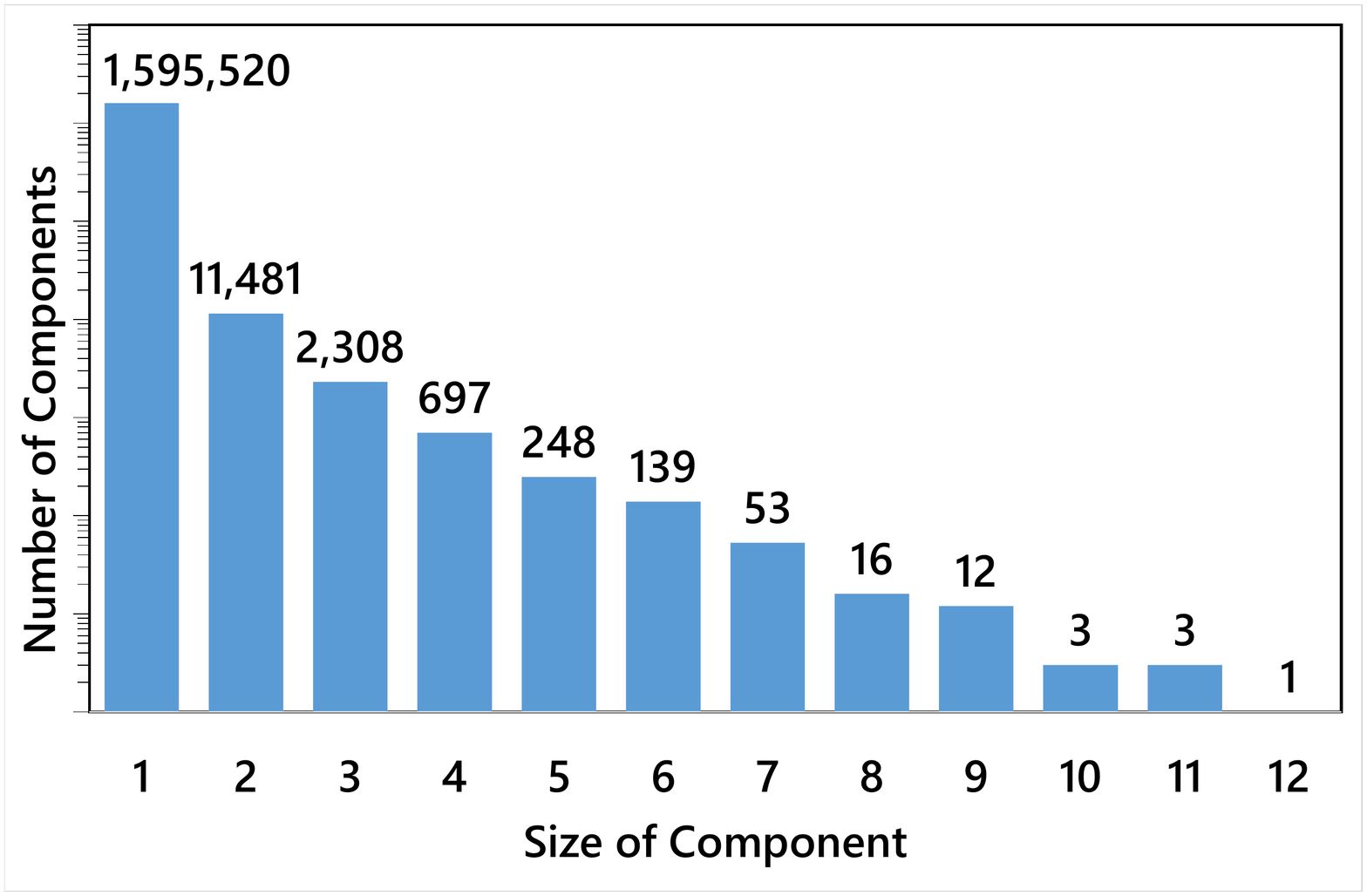}
  \caption{Statistics of strongly connected components sizes in VG.
Note that the y axis is logarithmic.}
  \label{fig:scc-sizes}
\end{figure}

From the results, we could see that CIR is negatively correlated to the compression ratio. The proposed method performs closely to the brute-force solver in CIR, therefore achieving a near-optimal compression ratio. The main reason is that the sizes of most strongly connected components are small, which is illustrated in \autoref{fig:scc-sizes}.
Since above 99\% nodes in VG are single components (\autoref{fig:scc-sizes}), they could be successfully sorted in topological order by the proposed preprocessor.

\subsubsection{Learned Distribution Model}
\label{sec:experiment-ablation-cdm}
This part tries to prove that the proposed learned distribution model contributes to numerical data compression under complicated distribution constraints in scene graphs. Specifically, we apply different distribution models for object location compression, including the Gaussian model, Laplacian model, Gaussian mixture model with 5 mixtures (GMM5), Gaussian mixture model with 10 mixtures  (GMM10), the proposed fully-dynamic variant model (FullDyn), and the proposed two-part model (Ours (LearnDist)). Note that the fully-dynamic variant contains two layers (as opposed to four layers in the two-part model), simulating the removal of the two trainable layers in the proposed two-part model.
For simplicity, only object location is compressed by the proposed framework in those experiments, while the element predictor for object location still utilizes other data elements. Therefore, only the compression ratio of object location is reported. 
The results are summarized in \autoref{tab:abl-cdm}.

\begin{table}[!t]
    \centering
    \scriptsize
    \setlength{\tabcolsep}{4pt}
    \caption{Compression ratio of object location using different distribution models.}
    \begin{tabular}{c|cc}
    \hline
    \diagbox{Model}{Dataset} & VG & VG-Gen \\
    \hline
    Gaussian & 29.95\% & 28.29\% \\
    Laplacian & 29.87\% (-0.28\%) & 28.51\% (0.76\%) \\
    GMM5 & 29.11\% (-2.82\%) & 27.62\% (-2.39\%) \\
    GMM10 & 29.04\% (-3.04\%) & 27.56\% (-2.59\%) \\
    FullDyn & 29.12\% (-2.78\%) & 27.46\% (-2.95\%) \\
    Ours (LearnDist) & \textbf{28.98\% (-3.24\%)} & \textbf{27.40\% (-3.16\%)} \\
    \hline
    \diagbox{Model}{Dataset} & VG-GenRel & HiEve \\
    \hline
    Gaussian & 34.08\% & 35.17\% \\
    Laplacian & 34.00\% (-0.24\%) & 36.00\% (2.37\%) \\
    GMM5 & 33.32\% (-2.22\%) & 35.25\% (0.25\%) \\
    GMM10 & 33.00\% (-3.18\%) & 35.77\% (1.72\%) \\
    FullDyn & 33.55\% (-1.55\%) & 41.34\% (17.55\%) \\
    Ours (LearnDist) & \textbf{32.89\% (-3.49\%)} & \textbf{35.02\% (-0.41\%)} \\
    \hline
    \end{tabular}
    \label{tab:abl-cdm}
\end{table}

Generally, the complex models (GMM and the proposed models) with more distribution parameters perform better than simple models (Gaussian and Laplacian). Our proposed two-part learned model owns a slight compression ratio than the Gaussian mixture model in all datasets.
As for our proposed two-part variant and the fully-dynamic variant for the distribution model, both perform similarly on VG-based datasets. However, the fully-dynamic one does not achieve ideal performance on HiEve, presumably caused by the optimization issue
discussed in \autoref{sec:method-dm}.
Therefore, it is preferred to use the two-part variant for more stable performance.
Moreover, on HiEve, all complex models except for the proposed two-part model perform worse than the simple Gaussian model, possibly because the HiEve graph only indicates the same action between two objects and does not guarantee position constraints. As most complex models are harder to optimize, they fail to model simple distributions in such cases. The proposed model contains trainable layers to help optimization under simple distributions cases, therefore achieving similar performance to the simple Gaussian model.

\begin{figure}[!t]
  \centering
  \subfloat[]{\includegraphics[width=0.16\textwidth]{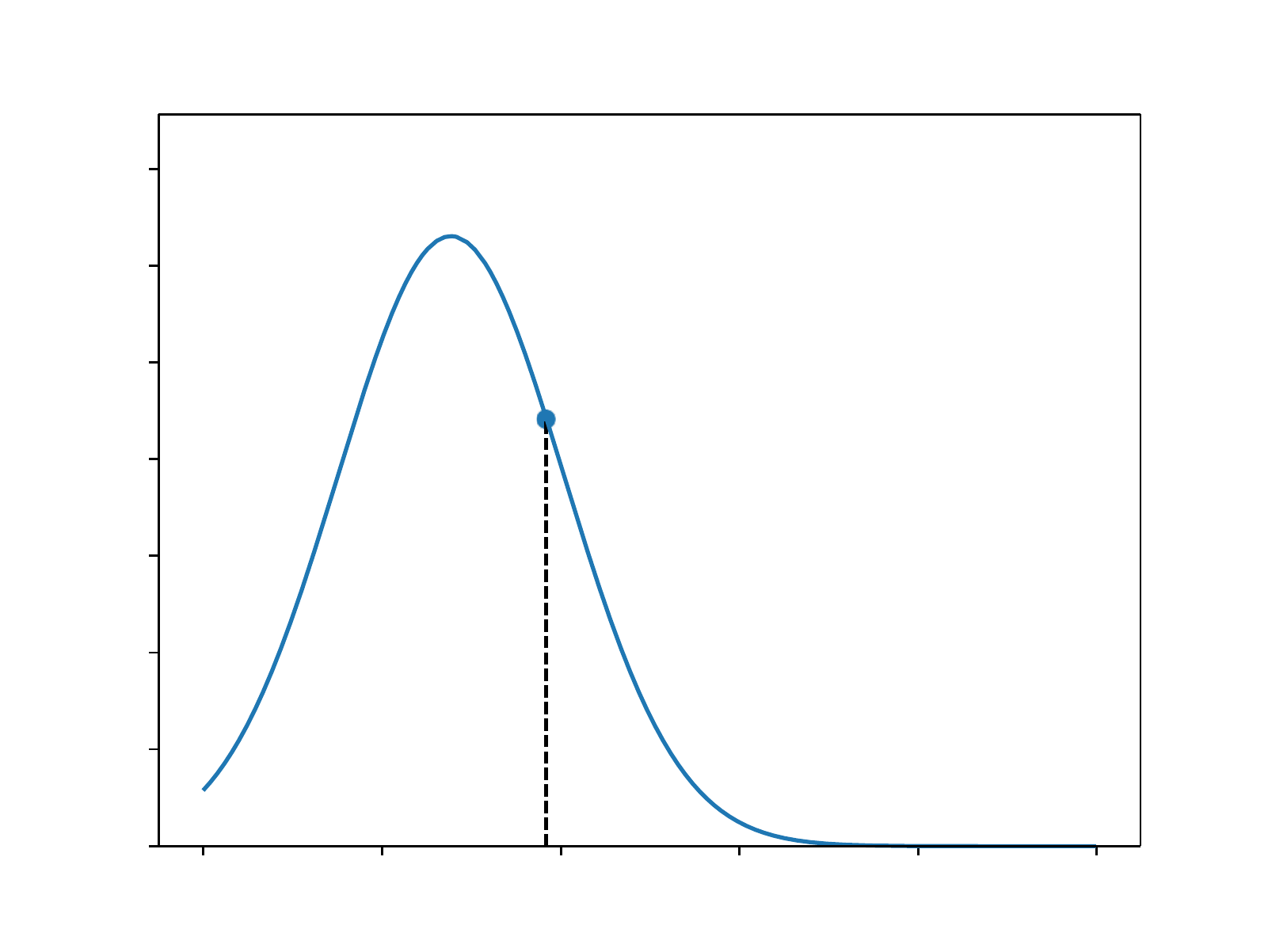}
  \label{fig:dist-gaussian}}
  \subfloat[]{\includegraphics[width=0.16\textwidth]{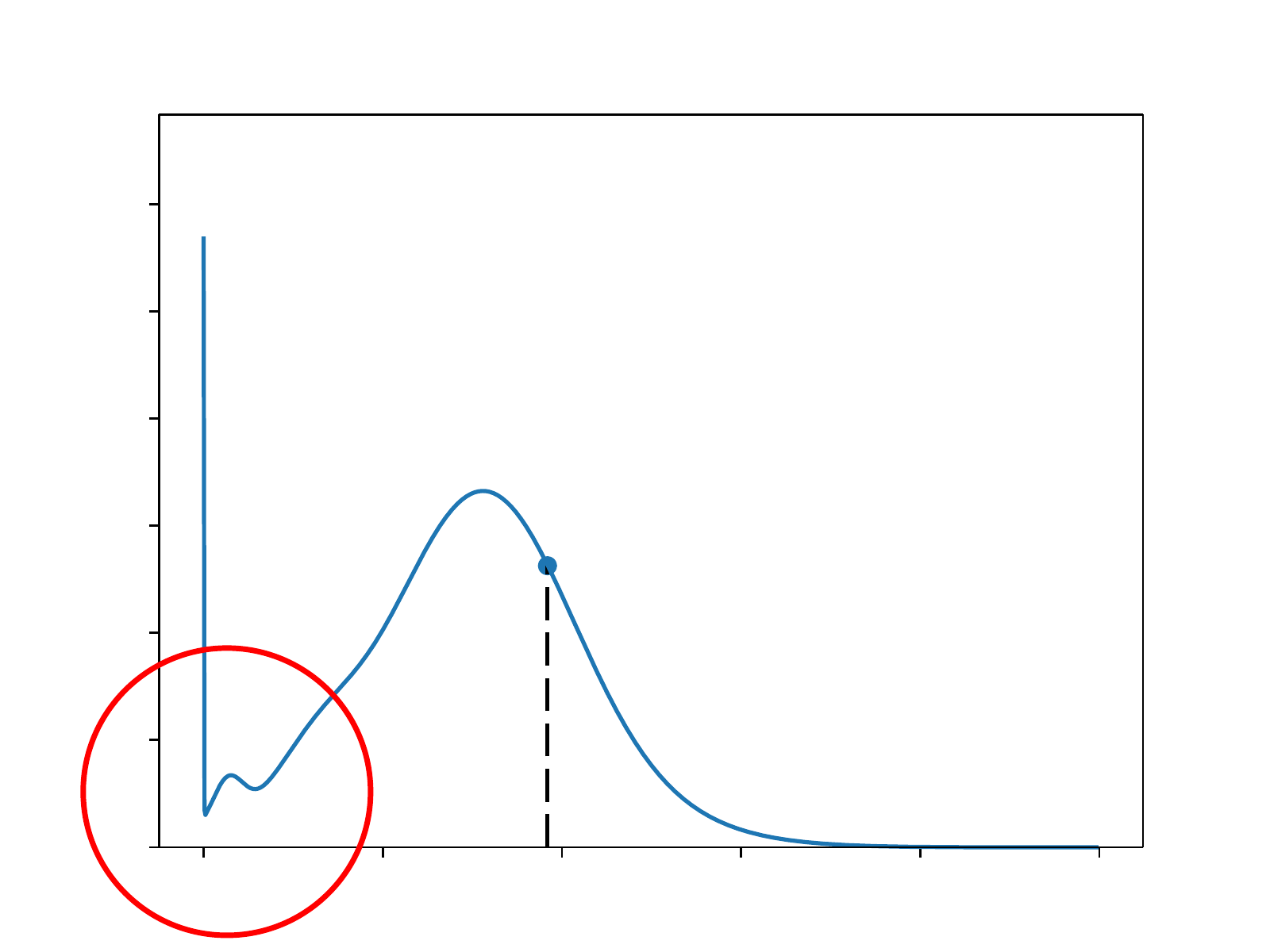}
  \label{fig:dist-gmm}}
  \subfloat[]{\includegraphics[width=0.16\textwidth]{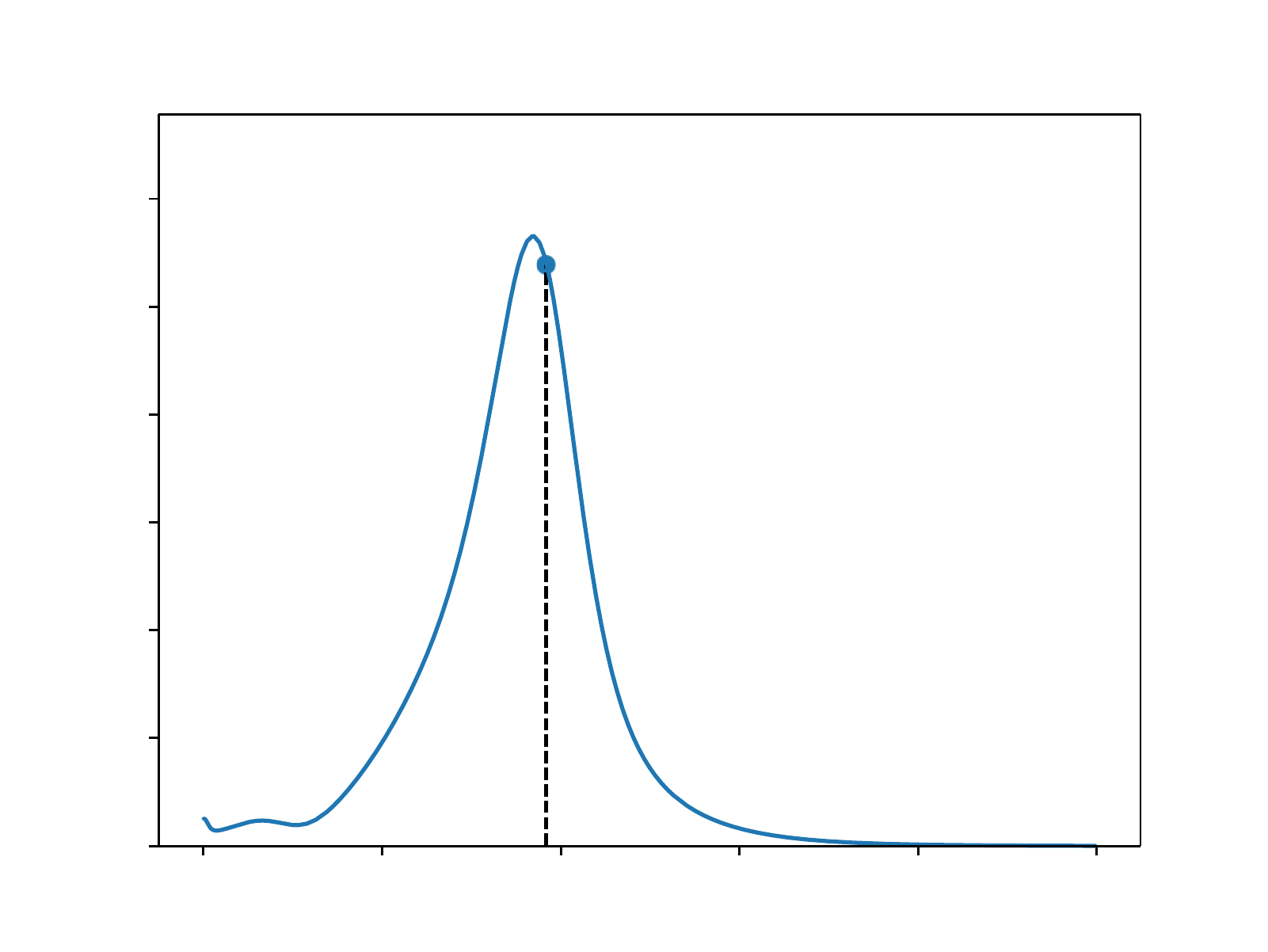}
  \label{fig:dist-gcd}}
  \caption{Probability distribution function visualization. 
  Comparison between: (a) Gaussian Distribution Model. (b) Gaussian Mixture Distribution Model (5 mixtures). (c) The proposed Learned  Distribution Model. 
  }
  \label{fig:dist-showcase}
\end{figure}

Furthermore, to explicitly present the improvement of the proposed learned distribution model, a \emph{qualitative experiment} is conducted to show the predicted probability distribution function of different distribution models in our framework. 
The results are shown in \autoref{fig:dist-showcase}. Apparently, the Gaussian model cannot cut the distribution on the value boundary (0 in the object location case as objects cannot exceed image boundary). 
Moreover, the Gaussian model is always symmetric along its mean value. This feature makes it unable to fit complicated conditional distribution cases well. The Gaussian mixture model performs better in fitting hard edges and complicated conditional cases, but the result still shows curves of single Gaussian modals, as shown in the red circle in \autoref{fig:dist-showcase}\redhighlight{b}, which is not the best fit result.
The proposed learned model shows similar fit result with Gaussian mixture model, meanwhile being more capable of fitting value boundary and conditional cases, showing asymmetric and confidential result.

\subsection{Effect of the Hyperparameters}
The proposed framework contains a few hyperparameters that can be adjusted for different datasets to achieve better compression results. The hyperparameter set mainly includes the unified prior extractor model parameters and the compression order of the element predictors.
A series of ablation studies are conducted to reveal the effect of different hyperparameters on our proposed framework.

\subsubsection{Unified Prior Extractor Implementation}
\label{sec:abl-pcm}
The unified prior extractor extracts vital prior information from scene graphs. Therefore, the implementation of this model also affects the performance of the compression framework. 
The backbone network structure of the unified prior extractor controls how to extract information from the original graph, while the width of the latent vector controls how much information is extracted from the original graph. 
Consequently, different backbone network operations, the number of layers, and the number of latent channels are explored in this experiment. 
Specifically, for backbone network operations, we compare Fully-Connected layers (FC), Graph Convolution Network (GCN)\cite{DBLP:conf/iclr/KipfW17} and Jumping-Knowledge Network (JKNet)\cite{DBLP:conf/icml/XuLTSKJ18} for the encoder.
The experiments are conducted on the VG dataset  (mainly composed of node data) and the VG-GenRel dataset (mainly composed of edge data). The results are shown in \autoref{tab:abl-pcm}.

\begin{table}[!t]
  \centering
  \caption{Compression Ratio using different unified prior extractor settings.}
  \label{tab:abl-pcm}
  \begin{minipage}{1.\linewidth}
    \centering
    \setlength{\tabcolsep}{2pt}
    \begin{tabular}{ccc|cc}
      \hline
      Backbone & Layer & Ch & VG & VG-GenRel \\
      \hline
      FC & 4 & 2 & \textbf{30.19\%} & 16.99\%  \\
    FC & 4 & 4 & 30.56\% (1.23\%) & 16.87\% (-0.73\%) \\
    FC & 8 & 2 & 30.21\% (0.08\%) & 17.05\% (0.38\%) \\
    FC & 8 & 4 & 30.72\% (1.77\%) & \textbf{16.81\% (-1.08\%)} \\
    GCN\cite{DBLP:conf/iclr/KipfW17} & 4 & 2 & 30.56\% (1.24\%) & 17.13\% (0.83\%) \\
    JKNet\cite{DBLP:conf/icml/XuLTSKJ18} & 4 & 2 & 30.24\% (0.18\%) & 17.00\% (0.08\%) \\
      \hline
    \end{tabular}
  \end{minipage}
\end{table}

\textbf{Backbone Network Operations}:
The results shows that the convolution-based backbones cannot further improve the performance.
Convolution-based backbones could improve structure compression to some extent by embedding structure information into the latent vector. Nevertheless, information for node data in the prior information is occupied by structure information. It results in a worse compression ratio for node data. Therefore, the overall result of convolution-based backbones underperforms the FC backbone, which is why we choose the simple FC backbone in our proposed framework.
Note that JKNet, which performs better in graph classification than GCN, also provides better results in scene graph compression than GCN. This phenomenon indicates that there might be better choices of operations by applying better convolution-based networks.

\textbf{Number of Layers}:
The results shows that the proposed framework cannot gain performance improvement by increasing the number of layers. The reason for this is that the FC backbone only extracts the information of node data, which is a relatively simple task and using complex models may increase the difficulty of optimization.

\textbf{Number of Latent Channels}:
Increasing the latent channels generally results in a better compression for data elements, while making the unified prior bitstream grow. This is a trade-off that may perform differently in different datasets. 
From the results, it could be found that increasing latent channels may reduce the total compression ratio in VG, while in VG-GenRel, increasing latent channels causes a slightly better compression ratio. Therefore, we could adjust latent channels for different datasets in order to get a better compression ratio, while the difference is generally not significant.

\subsubsection{Compression Order}
\begin{figure*}[!t]
  \centering
  \subfloat[]{\includegraphics[width=0.33\textwidth]{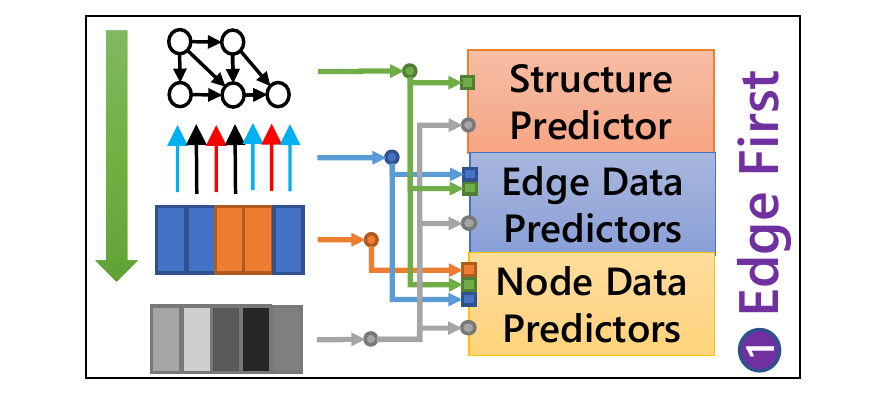}
  \label{fig:comp-orders-ef}}
  \subfloat[]{\includegraphics[width=0.33\textwidth]{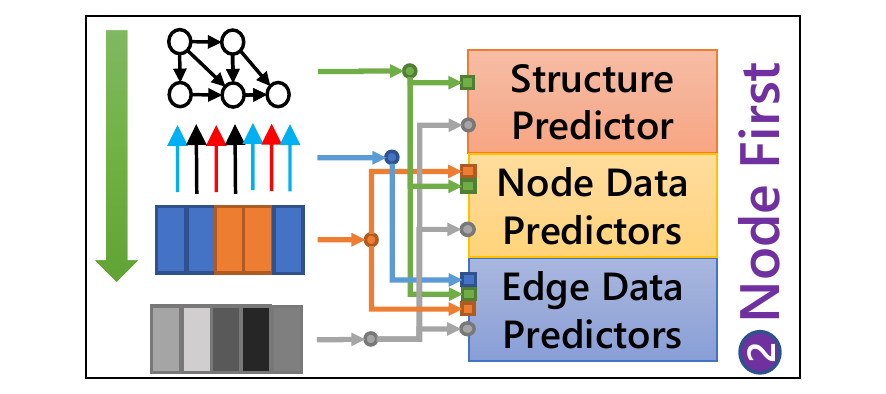}
  \label{fig:comp-orders-nf}}
  \subfloat[]{\includegraphics[width=0.33\textwidth]{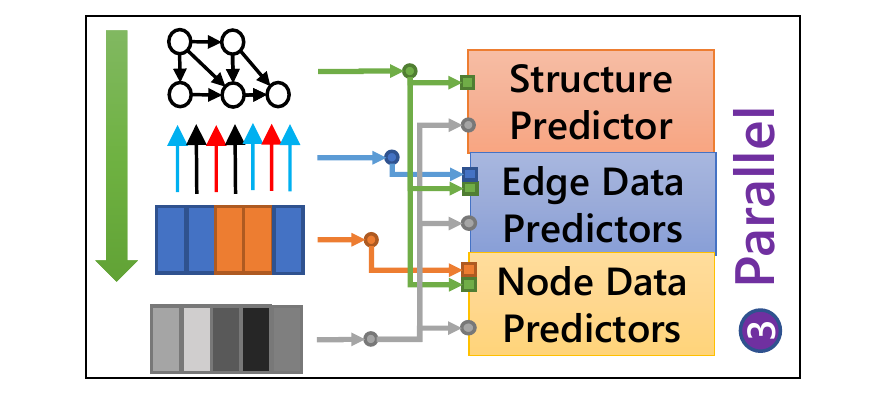}
  \label{fig:comp-orders-pl}}
  \caption{Implementation diagram of element predictors for different compression orders: (a) Edge First. (b) Node First. (c) Parallel. 
  }
  \label{fig:comp-orders}
\end{figure*}

The compression order of the graph elements in the scene graph affects how the context information is exploited in the element predictors, thus impacting the final compression ratio. 
Since graph structure is required for all context models, only the order of node data and edge data could be adjusted. Therefore, there are basically three options for compression order, the "Edge First" order, the "Node First" order, and the "Parallel" order, as illustrated in \autoref{fig:comp-orders}. 
For example, if edge data is decompressed first, then the node data predictor could utilize decompressed edge data by applying RGCN-based\cite{DBLP:conf/esws/SchlichtkrullKB18} context model to perform better prediction, as illustrated in \autoref{fig:comp-orders}\redhighlight{a}. 

\begin{figure}
  \includegraphics[width=1\linewidth]{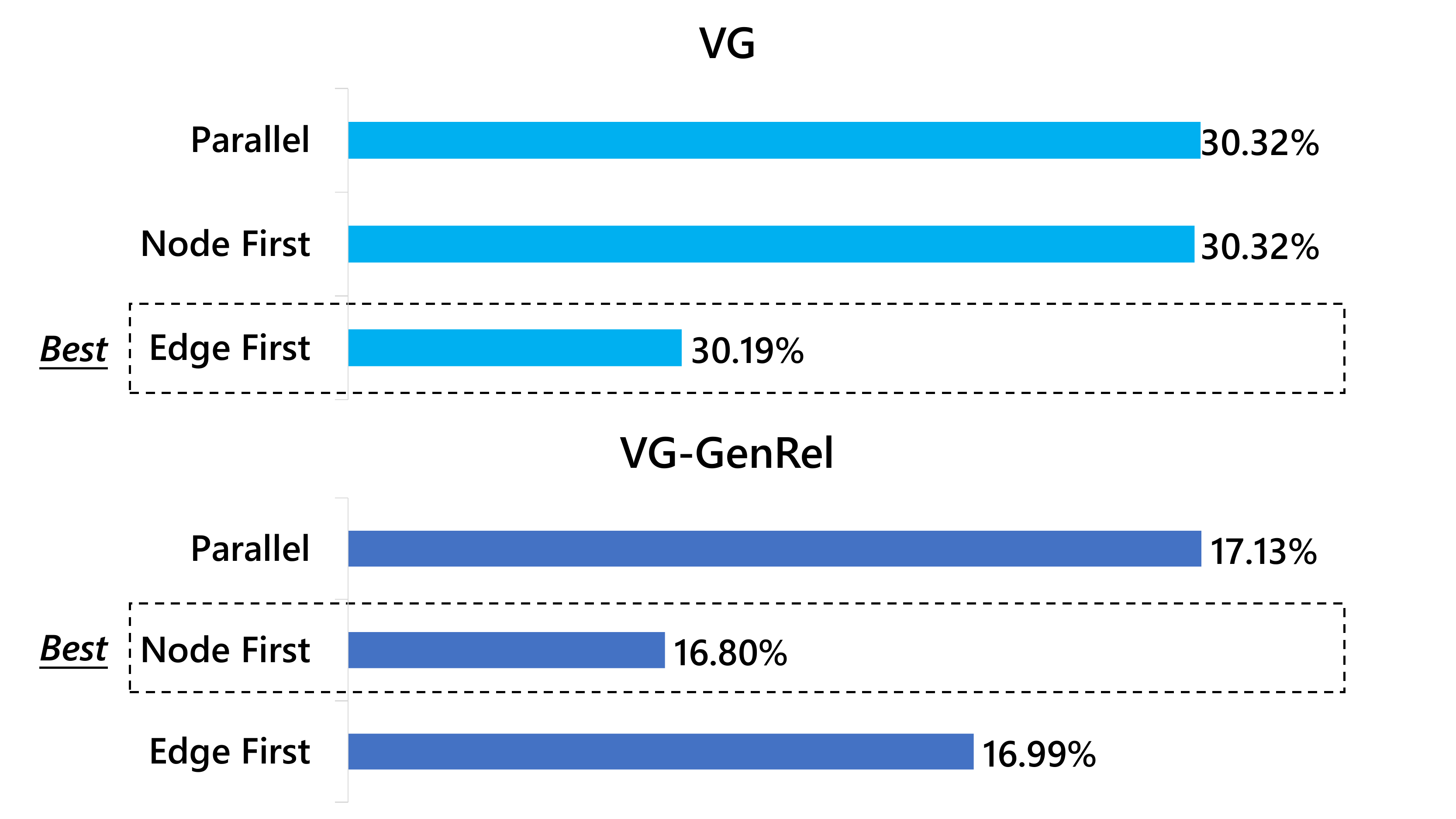}
  \caption{Compression ratio for different compression orders in different datasets. 
  }
  \label{fig:abl-ds}
\end{figure}

Consequently, different compression order is explored to form different structures of element predictors. The node-major dataset VG, as well as the edge-major dataset VG-GenRel, are considered for this experiment. The results are shown in \autoref{fig:abl-ds}. 

It could be found from the results that for the VG dataset, the `Edge First' order provides a slightly better result than other orders. Since the majority of data in VG is node data, the node data predictor could benefit more from former edge data for prediction to reduce the total entropy using `Edge First' order. Moreover, for VG-GenRel, `Node First' performs better as VG-GenRel data are mainly edge data. However, although different compression orders could be selected to better compress different datasets, the difference is generally not significant.

\section{Conclusion}
\label{sec:conclusion}
This paper mainly introduces a novel framework for joint lossless compression of the scene graph data. The main contributions include: (1) A adaptive compression framework with element-adaptive predictors for different data elements. (2) A series of graph context modeling approaches for different graph elements to exploit context redundancy. (3) A learned distribution model to model the distribution of graph-constrained numerical data elements for better prediction.
Experiments show that the proposed system outperforms state-of-the-art general-purpose codecs and combined codecs for total scene graph lossless compression, and also keep up with traditional graph structure compression methods in graph structure compression.


\section*{Declarations}
\subsection*{Data Availability Statement}
The datasets generated during and/or analysed during the current study are all available publicly. Please refer to \autoref{sec:datasets} for further details.

%
%

\bibliographystyle{spmpsci}      
\bibliography{IEEEabrv,egbib}   


\end{document}